\documentclass[acmsmall]{acmart}
\usepackage{xcolor}
\usepackage{multirow}
\usepackage[linesnumbered,ruled,vlined]{algorithm2e}
\AtBeginDocument{%
  }

\setcopyright{acmlicensed}
\copyrightyear{2025}
\acmYear{2025}
\acmDOI{XXXXXXX.XXXXXXX}

\acmJournal{JACM}
\acmVolume{1}
\acmNumber{1}
\acmArticle{111}
\acmMonth{1}

\begin{document}

\title{Divide-and-Conquer: Cold-Start Bundle Recommendation via Mixture of Diffusion Experts}


\author{Ming Li}
\email{liming7677@whut.edu.cn}
\affiliation{
  \institution{York University}
  \city{Toronto}
  \country{Canada}
}
\affiliation{
  \institution{Wuhan University of Technology}
  \city{Wuhan}
  \country{China}
}

\author{Lin Li}
\email{cathylilin@whut.edu.cn}
\authornote{Corresponding author}
\affiliation{
  \institution{Wuhan University of Technology}
  \city{Wuhan}
  \country{China}
}

\author{Xiaohui Tao}
\email{Xiaohui.Tao@unisq.edu.au}
\affiliation{
  \institution{University of Southern Queensland}
  \city{Toowoomba}
  \country{Australia}
}

\author{Dong Zhang}
\email{zhangdong23@whut.edu.cn}
\affiliation{
  \institution{Wuhan University of Technology}
  \city{Wuhan}
  \country{China}
}

\author{Jimmy Xiangji Huang}
\email{jhuang@yorku.ca}
\affiliation{
  \institution{York University}
  \city{Toronto}
  \country{Canada}
}

\renewcommand{\shortauthors}{Li et al.}

\begin{abstract}
Cold-start bundle recommendation focuses on modeling new bundles with insufficient information to provide recommendations. 
Advanced bundle recommendation models usually learn bundle representations from multiple views (e.g., interaction view) at both the bundle and item levels. 
Consequently, the cold-start problem for bundles is more challenging than that for traditional items due to the dual-level multi-view complexity. 
In this paper, we propose a novel \textbf{M}ixture of \textbf{Diff}usion \textbf{E}xperts (\textbf{MoDiffE}) framework, which employs a divide-and-conquer strategy for cold-start bundle recommendation and follows three steps:
(1) \textbf{Divide}: The bundle cold-start problem is divided into independent but similar sub-problems sequentially by level and view, which can be summarized as the poor representation of feature-missing bundles in prior-embedding models. 
(2) \textbf{Conquer}: Beyond prior-embedding models that fundamentally provide the embedded representations, 
we introduce a diffusion-based method to solve all sub-problems in a unified way, which directly generates diffusion representations using diffusion models without depending on specific features.  
(3) \textbf{Combine}: A cold-aware hierarchical Mixture of Experts (MoE) is employed to combine results of the sub-problems for final recommendations, where the two models for each view serve as experts and are adaptively fused for different bundles in a multi-layer manner. 
Additionally, MoDiffE adopts a multi-stage decoupled training pipeline and introduces a cold-start gating augmentation method to enable the training of gating for cold bundles.
Through extensive experiments on three real-world datasets, we demonstrate that MoDiffE significantly outperforms existing solutions in handling cold-start bundle recommendation. It achieves up to a 0.1027 absolute gain in Recall@20 in cold-start scenarios and up to a 47.43\% relative improvement in all-bundle scenarios.

\end{abstract}

\ccsdesc[500]{Information systems~Information retrieval}

\keywords{Cold-Start Bundle Recommendation, Divide-and-Conquer, Diffusion Model, Mixture of Experts}


\maketitle

\section{Introduction}
As a powerful tool to alleviate information overload, recommendation systems have succeeded greatly and become an indispensable part of modern life. 
Unlike traditional recommendation systems that recommend a single item, bundle recommendation aims to recommend a bundle of items designed to be consumed as a whole~\cite{bundle_rec_2014a,bundle_rec_2014b,DAM,HyperMBR,MultiCBR,CrossCBR,DSCBR,MIDGN}. 
In recent years, bundle recommendations have continued to receive great attention from industry and academia due to their ability to enhance user experience and assist merchants in increasing sales~\cite{bundle_survey_2020, bundle_survey_2024}. 
These systems have found successful applications across various domains, such as e-commerce bundle~\cite{revisite_bundle_2022,revisite_bundle_2024,bundle_rec_2014a,bundle_rec_2014b}, music playlist~\cite{BGCN,AttList,MIDGN,HyperMBR}, meal~\cite{mealrec+, healthy_meal_rec,CateRec}, fashion outfit~\cite{outfit_2019,CrossCBR,MultiCBR,DSCBR}, demonstrating their versatility and effectiveness.

\begin{figure}[h]
\includegraphics[width=\textwidth]{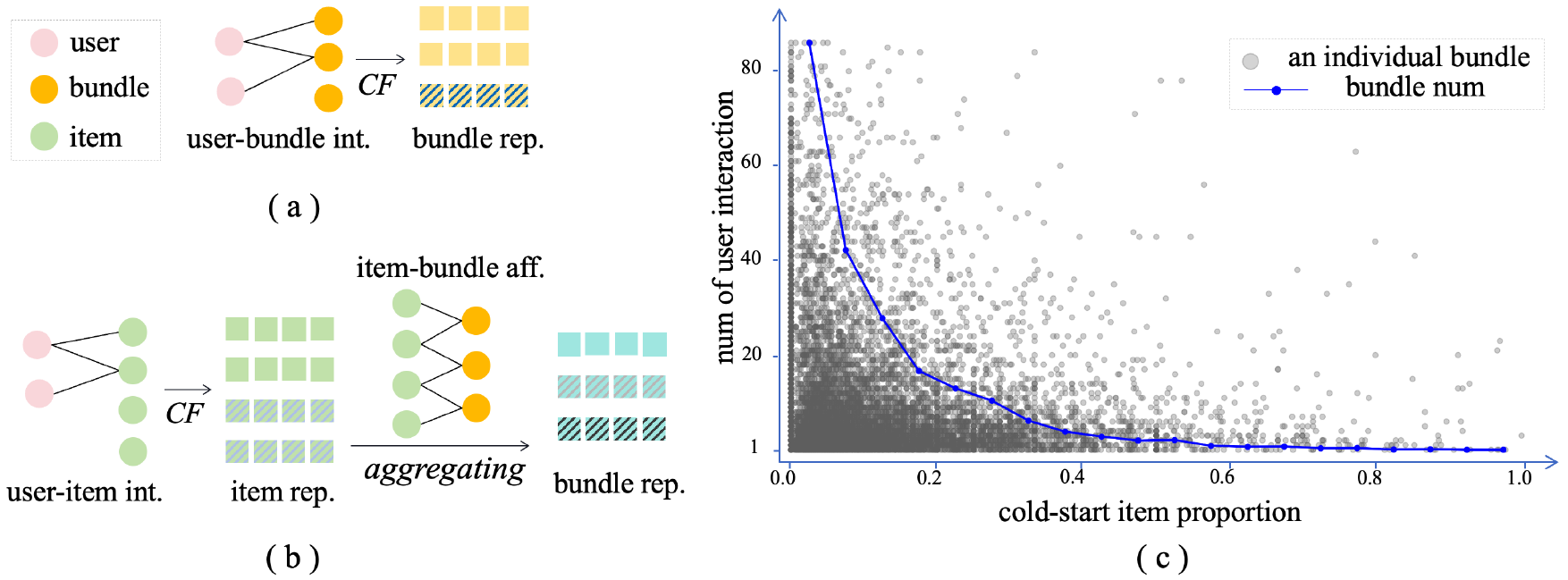}
\caption{(a) Bundle-level interaction-view representation learning. (b) Item-level interaction-view representation learning. (c) Bundle distribution in the real-world dataset NetEase. }
\label{cold-example}
\end{figure}

The non-atomic nature of bundles is a key characteristic in bundle recommendation, meaning bundles contain information at two distinct levels of granularity: the bundle level and the item level. Each level contains multiple views of information. Therefore, \textbf{the bundle recommendation task has dual-level multi-view complexity. }
A bundle can be represented by multiple views of its own bundle-level information, as well as multiple views of the item-level information it contains.
Existing research on bundle recommendation~\cite{BGCN,CrossCBR,HyperMBR,CateRec,MultiCBR,DSCBR} gradually evolves into a dual-level multi-view architecture. Taking the important interaction information as an example, as illustrated in Figure~\ref{cold-example}(a) and (b), the bundle-level interaction view directly learns the bundle representation from the historical user-bundle interactions. In contrast, the item-level interaction view learns item representations from the user-item interactions and then aggregates these item representations into bundle representations according to bundle-item affiliations. 
Representations from different views capture knowledge about users and bundles under different perspectives, complementing each other and proving to be indispensable. 

The cold-start problem is a fundamental challenge in recommender systems, where the model struggles to learn effectively and provide recommendations for new users and items with insufficient information~\cite{cold_start_survey_2017,cold_start_survey_2025}. 
Addressing the cold-start problem in bundle recommendation is crucial to ensure constantly emerging new bundles gain visibility, deliver immediately relevant suggestions that build user trust and loyalty, and drive discoverability and revenue growth. 
However, the cold-start problem for bundles is more challenging than that for traditional items, as the bundle recommendation model learns bundle representations from multiple views of dual levels. 
\textbf{We are the first to emphasize the dual-level multi-view complexity of the bundle cold-start problem.} 
The cold-start problems of different views at different levels collectively form the bundle cold-start problem. 
From both model and data perspectives, the bundle cold-start problem presents two main characteristics that pose distinct challenges: 

\textbf{(1) Different views at different levels have different cold-start problems.} Different views (such as interaction view and content view) utilize various representation learning models, and each level focuses on different entities (i.e., bundle and item).  
Taking the interaction view as an example, Figure~\ref{cold-example}(a) shows that the bundle-level interaction-view cold-start problem is that the model cannot effectively model bundles that lack historical user interactions, resulting in poor representation. 
Figure~\ref{cold-example}(b) illustrates the item-level interaction-view cold-start problem, which aggregates the poorly represented cold item representations into the bundle representation, negatively impacting the representation quality. 
Given that different views are heterogeneous but indispensable, \textbf{simultaneously addressing cold-start problems of multiple views at dual levels poses a significant challenge.}

\textbf{(2) Diverse cold-start situations of bundles coexist.} 
Figure~\ref{cold-example} (c) illustrates the distribution of bundles in terms of interaction information in the real-world dataset from NetEase. 
It reveals that a substantial portion of the bundles contain cold items, and the proportion of cold items exhibits a smooth distribution. 
Moreover, the less popular the bundle, the higher the proportion of cold items it contains, and the more serious the item-level interaction-view cold-start problem is. 
Therefore, cold-start problems in different views and levels are often intertwined, resulting in diverse cold-start situations coexisting in one domain. 
For instance, in a real-world e-commerce arena, a bundle might be a promoted combination of existing products, a value-added bundle containing unsalable items, or a newly launched product collections, etc.
\textbf{This diversity poses a significant challenge for models to adaptive handle for diverse cold-start bundle situations. }

\textbf{However, existing methods have not effectively solved the bundle cold-start problem.} (1) An intuitive solution is to treat bundles as atomic units and apply the existing item-based cold-start methods~\cite{item_cold_start_CB2CF_2019,item_cold_start_CCFCRec_2023,item_cold_start_CVAR_2022,item_cold_start_DropoutNet_2017,item_cold_start_GAR_2022,item_cold_start_Heater_2020a,item_cold_start_CLCRec_2021} to solve the bundle cold-start problem. 
However, these methods can only solve the bundle-level problem but ignore the item-level problem, resulting in suboptimal recommendation performance. (2) Existing research on cold-start bundle recommendation~\cite{bundle_cold_start_CoHeat_2024} relies on item-level information of cold bundle. However, they fail to realize the item-level cold-start problem that cold items exist in cold bundles. Although they alleviate the bundle cold-start problem by emphasizing item-level representation learning, relying solely on existing item-level information is insufficient to represent bundles fully. 

\textbf{To address the bundle cold-start problem, we present MoDiffE: a novel Mixture of Diffusion Experts framework that employs a divide-and-conquer strategy powered by diffusion models and MoE. }This framework follows three steps: 
\begin{itemize}
    \item \textbf{Divide}: The bundle cold-start problem is divided into independent yet similar sub-problems sequentially by level and view. This division simplifies the overall complexity by breaking the problem into more manageable yet similar sub-problems. 
    These sub-problems are summarized as a unified representation learning challenge: the poor representation of feature-missing bundles in prior-embedding models. 
    \item \textbf{Conquer}: A diffusion-based method is proposed to solve all sub-problems in a unified way. 
    Specifically, beyond prior-embedding models that fundamentally provide the embedded representations, diffusion models are trained to capture the underlying distribution of well-represented representations through a diffusion-denoise process, enabling the direct generation of high-quality diffusion representations without relying on specific features. 
    This method bypasses the differences in representation learning across different levels and views, solving the first challenge. 
    \item \textbf{Combine}: A cold-aware hierarchical Mixture of Experts (MoE) architecture is proposed to combines results of individual sub-problems to form a solution to the original bundle cold-start problem. 
    Specifically, the two models for each view serve as experts. They are first competitively fused into view-specific results with cold-aware features and then cooperatively fused into the final recommendation. 
    This method improves the stability of the fusion process and the adaptability to diverse cold-start situations, solving the second challenge. 
\end{itemize}
MoDiffE adopts a multi-stage decoupled pipeline for training, enhancing the framework's flexibility and adaptability. 
In addition, since the cold-start bundle is is absent from the training set, the gating networks in MoE cannot handle it accurately during inference. 
To address this, we propose a cold-start gating augmentation method that synthesizes pseudo cold bundles through interpolation during training. 
Through extensive experiments on real-world datasets, we demonstrate that MoDiffE significantly outperforms state-of-the-art methods in cold-start bundle recommendation. 
Specifically, MoDiffE achieves up to a 0.1027 absolute gain in Recall@20 in the cold-start scenarios and up to a 47.43\% relative improvement in all-bundle scenarios. 
Our contributions can be summarized as follows:
\begin{itemize}
    \item We are the first to emphasize the dual-level multi-view complexity of the bundle cold-start problem, which poses challenges in simultaneously addressing cold-start problems of multiple views at dual levels and adaptive handling diverse cold-start situations of bundles. 
    \item We propose the Mixture of Diffusion Experts (MoDiffE) framework with a divide-and-conquer strategy, which leverages diffusion models to solve different cold-start sub-problems uniformly and uses a cold-aware hierarchical MoE to handle different bundles adaptively. 
    \item We conduct extensive experiments on three real-world datasets and demonstrate MoDiffE significantly outperforms state-of-the-art methods in cold-start bundle recommendation. 
\end{itemize}

\section{Related Work}

\subsection{Bundle Recommendation}
Bundle recommendation aims to enhance user satisfaction and drive sales by recommending a bundle of items designed to be consumed as a whole. 
Bundle recommendation is characterized by the non-atomic nature of bundle, where bundles contain information at two distinct levels: the bundle level and the item level.
Each level contains multiple views of information~\cite{bundle_survey_2020,bundle_survey_2024}.
The most concern for researchers is the interaction views at dual levels. 
The earlier studies mainly focus on user-bundle interactions and use Bayesian Personalized Ranking~\cite{bundle_rec_2014a,bundle_rec_2017a} and embedding factorization models~\cite{bundle_rec_2017b} for bundle recommendation. 
DAM~\cite{DAM} introduces user-item interactions and uses a multi-task framework to capture shared collaborative filtering signals between bundle and item interactions but does not effectively utilize bundle-item affiliation. 

With the rapid development of graph machine learning~\cite{LightGCN,graph_on_rec_survey_2023} in the field of recommendation systems, bundle recommendation researches~\cite{BGCN,MIDGN,CrossCBR,HyperMBR,CateRec,MultiCBR} have begun to use graph structures to model the complex relationships between users, bundles, and items. 
BGCN~\cite{BGCN} first applies Graph Convolutional Networks to bundle recommendation and proposes a dual-level multi-view architecture consisting of bundle-level interaction view and item-level interaction view, as shown in Figure~\ref{cold-example}(a) and (b). 
This dual-level multi-view architecture makes great progress in bundle recommendation, and subsequent research is mainly based on this architecture to improve. 
CrossCBR~\cite{CrossCBR} uses LightGCN~\cite{LightGCN} as the backbone model and utilizes contrastive learning between two views to capture cooperative association signals. 
HyperMBR~\cite{HyperMBR} encodes two views interaction graphs in hyperbolic space to learn accurate hyperbolic representations, and proposes a hyperbolic distance-based mutual distillation to encourage the two views to transfer knowledge. 
AMCBR~\cite{AMCBR} and MultiCBR~\cite{MultiCBR} build additional graphs and hypergraphs on two-view interaction graphs and use multi-view contrastive learning to enhance representation. 

\subsection{Cold-Start Recommendation}

\subsubsection{Cold-Start Item Recommendation}
Cold-start problem is one of the long-standing challenges in recommender systems, focusing on accurately modeling new users or items with insufficient information to provide recommendations~\cite{cold_start_survey_2017,cold_start_survey_2025}.
Existing approaches to address cold-start problems can be summarized in the following categories. 
Content-based methods~\cite{item_cold_start_content_2016,item_cold_start_content_2024} leverage auxiliary information (such as item attributes, user demographics, and textual descriptions) of items or users to build content-based representations to handle cold-start scenarios. 
Robustness-based methods~\cite{item_cold_start_dropout_2019,item_cold_start_dropout_2020,item_cold_start_DropoutNet_2017} randomly mask or drop out features during training, forcing the model to learn robust representations that generalize well to new items or users. 
Constraint-based methods~\cite{item_cold_start_CCFCRec_2023,item_cold_start_CB2CF_2019,item_cold_start_Heater_2020a,item_cold_start_CLCRec_2021} explicitly model the relationships between item content and collaborative embeddings by applying a constraint loss, where CLCRec~\cite{item_cold_start_CLCRec_2021} and CCFCRec~\cite{item_cold_start_CCFCRec_2023} are based on contrastive learning. 
Generative methods~\cite{item_cold_start_CVAR_2022,item_cold_start_GAR_2022,item_cold_start_generative_2019,item_cold_start_generative_2020} use generative models to learn the mapping between features and collaborative filtering signals, thereby directly generating representations using the features of cold items. For example, GAR~\cite{item_cold_start_GAR_2022} is based on the generative adversarial network, and CVAR~\cite{item_cold_start_CVAR_2022} is based on a conditional variational autoencoder. 
Meta-learning methods~\cite{item_cold_start_meta_learning_2022,item_cold_start_meta_learning_2023,item_cold_start_meta_learning_2021a,item_cold_start_meta_learning_2021b} enable models to learn user preferences globally in the pre-training stage followed by local fine-tuning for a target user with only a few interactions. 

\subsubsection{Cold-Start Bundle Recommendation}

Addressing the bundle cold-start problem is critical for bundle recommendation because it ensures newly created bundles without interaction history gain visibility in users’ feeds, enables immediate personalized suggestions that foster trust and long-term engagement, improves the diversity and discoverability of offerings, and ultimately boosts conversion rates and revenue growth for businesses. 
However, the cold-start problem for bundles is more challenging than that for traditional items, because bundle representations are derived from multiple views at dual levels, each of which suffers from distinct cold-start problems. 

Treating bundles as atomic units, the existing item-based cold-start method can be directly applied to bundle recommendation. 
However, these methods focus solely on addressing the bundle-level cold-start problem while overlooking the item-level cold-start problem that cold items exist in bundles, resulting in suboptimal recommendation performance. 

Hyunsik Jeon et al.~\cite{bundle_cold_start_CoHeat_2024} are the first to explore the bundle cold-start problem and introduced a framework named CoHeat. They emphasize the crucial role of user-item interactions in alleviating the bundle cold-start problem. 
Specifically, CoHeat designs a curriculum strategy that progressively transitions the training focus from user-bundle interactions to user-item interactions. This enables the model to infer user preferences for new bundles based on user-item interactions. 
CoHeat is a solid step forward in studying cold-start bundle recommendation, but it still has some limitations. 
First, CoHeat does not realize the dual-level complexity of bundle cold-start problem and only focuses on the bundle-level cold-start problem. 
Second, relying only on existing user-item interactions is not enough to fully represent cold bundles. 
On the one hand, bundle-level and item-level representation are different, complementary, and indispensable. 
On the other hand, aggregating underrepresented cold item representations compromises the quality of the item-level bundle representation, while relying solely on warm items creates a compositional gap with the original bundle. 

We are first to emphasize the dual-level multi-view complexity of the bundle cold-start problem, which has not been effectively solved in existing works. 
In this paper, we formally define the bundle cold-start problem with dual-level multi-view complexity and propose a divide-and-conquer framework for solving it.

\section{Preliminary}
In this section, we first provide symbolic definitions of cold-start bundle recommendation. Then, we introduce the mainstream multi-view architecture in the current bundle recommendation. 

\subsection{Problem Definition}

The bundle cold-start problem arises due to insufficient information for accurately representing new bundles, which is further made more challenging by the dual-level multi-view complexity. 
The cold-start problems of different views at different levels collectively form the bundle
cold-start problem.
Without loss of generality, we consider the interaction view at dual levels while noting that other views can be analogously considered. 

Let $\mathcal{U}$, $\mathcal{B}$, and $\mathcal{I}$ denote the sets of users, bundles, and items, respectively. We follow the strict cold-start task setting in previous works~\cite{bundle_cold_start_CoHeat_2024,cold_start_survey_2025}. 
The definitions of cold bundle and bundle cold-start problem differ in different views. 
For simplicity, we use B-int and I-int to represent bundle-level and item-level interaction views, respectively. The dash before represents the level, and the after represents the view.

In the bundle-level interaction view, bundles are categorized into two subsets: $\mathcal{B}_{w}^{B\text{-}int} \subseteq \mathcal{B}$ refers to the B-int warm bundles that have at least one historical interaction with users, while $\mathcal{B}_{c}^{B\text{-}int} = \mathcal{B} \setminus \mathcal{B}_{w}^{B\text{-}int}$ represents the B-int cold bundles that lack any historical interaction with users. 
A challenge in the B-int representation learning for bundle recommendation models is that they cannot effectively represent cold bundles, resulting in the B-int cold-start problem. 

In the item-level interaction view, as the atomic entity, items are classified as warm items ($\mathcal{I}_{w}$) or cold items ($\mathcal{I}_{c}$) based on the availability of user interaction data. 
$\mathcal{C}_b$ refers to the item set of bundle $b \in \mathcal{B}$. $\mathcal{B}_{w}^{I-int} = \{ b \in \mathcal{B} \mid \mathcal{C}_b \cap \mathcal{I}_{c} = \varnothing \}$ refers to the I-int warm bundles that contains only warm items, while
$\mathcal{B}_{c}^{I-int} = \{ b \in \mathcal{B} \mid \mathcal{C}_b \cap \mathcal{I}_{c} \ne \varnothing \}$ refers to the I-int cold bundles that contain at least one cold item. 
The ratio of cold items within a bundle $b$ is represented as $r_b = |\mathcal{C}_b \cap \mathcal{I}_{\text{c}}|/{|\mathcal{C}_b|}$, which indicates the degree of cold start. 
For I-int representation learning of bundle recommendation models, they aggregate the poorly represented cold item representations into the I-int bundle representation, resulting in the I-int cold-start problem. 

Usually, the cold-start situation of the bundle in different views is independent of each other, and there are many possible combinations. Therefore, a bundle recommendation scenario contains a variety of cold-start situation bundles. 
This is reflected in the fact that the intersection of the following sets is usually non-empty:
\[
\mathcal{B}_{c}^{B\text{-}int} \cap \mathcal{B}_{c}^{I-int} \neq \varnothing, \quad
\mathcal{B}_{c}^{B\text{-}int} \cap \mathcal{B}_{w}^{I-int} \neq \varnothing, \quad
\mathcal{B}_{w}^{B\text{-}int} \cap \mathcal{B}_{c}^{I-int} \neq \varnothing, \quad
\mathcal{B}_{w}^{B\text{-}int} \cap \mathcal{B}_{w}^{I-int} \neq \varnothing.
\]

The cold-start bundle recommendation task is to recommend new bundles based on existing information.
The observed user-bundle interactions, user-item interactions, and bundle-item affiliations are defined respectively as $X = \{(u, b)\mid u \in \mathcal{U}, b \in \mathcal{B}_{w}^{B\text{-}int} \}$, $Y = \{(u, i)\mid u \in \mathcal{U}, i \in \mathcal{I}_w\}$, and $Z = \{(b, i)\mid b \in \mathcal{B}, i \in \mathcal{I}\}$. Given $\{X, Y, Z\}$, our goal is to recommend $k$ bundles from $\mathcal{B}$ to each user $u \in \mathcal{U}$. 
Note that the given interactions are observed only for B-int warm bundles, but the objective also includes recommending new (B-int cold) bundles to users. 
The term "cold bundles" in the following text refers specifically to bundle-level interaction-view cold bundles. Additionally, all bundles may contain cold items. 
When considering other views, such as content and knowledge graphs, the definition of the bundle cold-start problem for a specific view can also be analogously applied at dual levels.

The previous definition of bundle cold-start problem~\cite{bundle_cold_start_CoHeat_2024} does not realize the item-level cold-start problem. However, However, this definition contributes by explicitly emphasizes the dual-level multi-view complexity of the bundle cold-start problem. 

\subsection{Bundle Recommendation Model}\label{sec:Collaborative_Experts}
The non-atomic nature of bundles is a key characteristic in bundle recommendation, meaning bundles contain information at two distinct levels of granularity: bundle and item levels. 
Each level contains multiple views, such as interaction, content, knowledge graph, etc. 
Therefore, a bundle can be represented not only by multiple views of it own bundle-level information but also by multiple views of the item-level information it contains. 
Existing models usually use prior-embedding models to learn and represent bundle features. 
Taking the important interaction information as an example, this section explains the representation learning of interaction views at dual levels. The prior-embedding model under interaction view is usually collaborative filtering models.

\subsubsection{Bundle-Level Interaction View}
This view aims to learn collaborative filtering signals from user-bundle interactions. It can be formulated as follows: 
\[
R_{u}^{B-int,e}, R_{b}^{B-int,e} = \mathcal{CF}(X, E_{u}, E_{b})
\]
where $\mathcal{CF}(\cdot)$ refers to a collaborative filtering model and $R_{u}^{B-int,e} \in \mathbb{R}^{d'\times |\mathcal{U}|}$ and $R_{b}^{B-int,e} \in \mathbb{R}^{d'\times |\mathcal{U}|}$ are B-int embedded representations for users and bundles, respectively. $X$ is observed user-bundle interactions. $E_{u} \in \mathbb{R}^{d\times |\mathcal{U}|}$ and $E_{b} \in \mathbb{R}^{d\times |\mathcal{B}|}$ are the initial embeddings for users and bundles, respectively. $d$ and $d'$ are the embedding dimension and the representation dimension. 

$\mathcal{CF}(\cdot)$ can be implemented by any model, such as Matrix Factorization and Neural Collaborative Filtering~\cite{NCF_2017}. 
With the widespread application of Graph Machine Learning in recommendation systems, bundle recommendation researchers~\cite{BGCN,CrossCBR,MultiCBR,AMCBR,HyperMBR} are increasingly leveraging advanced graph neural network models. 
In this paper, we implement our MoDiffE framework with LightGCN~\cite{LightGCN} because it is simple but effective, and other methods are left as future work. 
Based on the user-bundle interaction matrix $X$, we construct a user-bundle bipartite graph and the graph propagation of bundle-level interaction view is depicted as: 
\begin{equation}
\begin{cases}\label{bundle_view_colaborative}
e_u^{B-int(k)}=\sum_{b\in\mathcal{N}_u}\frac{1}{\sqrt{|\mathcal{N}_u|}\sqrt{|\mathcal{N}_b|}}e_b^{B-int(k-1)}, \quad r_u^{B-int,e}=\textstyle \frac{1}{K}\sum_{k=0}^Ke_u^{B-int(k)} \\
e_b^{B-int(k)}=\sum_{u\in\mathcal{N}_b}\frac{1}{\sqrt{|\mathcal{N}_b|}\sqrt{|\mathcal{N}_u|}}e_u^{B-int(k-1)}, \quad r_b^{B-int,e}=\textstyle \frac{1}{K}\sum_{k=0}^Ke_b^{B-int(k)} & 
\end{cases}
\end{equation}
where $e_{u}^{B-int(k)}$, $e_{b}^{B-int(k)} \in \mathbb{R}^{d}$ are the $k$-th layer’s embeddings for user $u$ and bundle $b$; $\mathcal{N}_u$ and $\mathcal{N}_b$ are the neighbors of the user $u$ and bundle $b$ in the user-bundle graph. After obtaining embeddings from different layers, we pool them into B-int embedded representations $r_u^{B-int,e}$ and $r_b^{B-int,e}$. 

\subsubsection{Item-Level Interaction View}
This view learns I-int user and item embedded representations from user-item interactions, and then aggregates them into I-int bundle embedded representations based on the bundle-item affiliation. It can be formulated as follows: 
\begin{align}
    R_{u}^{I-int,e}, R_{i}^{I-int,e} &= \mathcal{CF}(Y, E_{u}, E_{i}) \\
R_{b}^{I-int,e} &= \mathcal{AGG}(R_{i}^{I-int,e}, Z)
\end{align}
where and $R_{u}^{I-int,e} \in \mathbb{R}^{d^{\prime}\times |\mathcal{U}|}$, $R_{i}^{I-int,e} \in \mathbb{R}^{d^{\prime}\times |\mathcal{I}|}$, and $R_{b}^{I-int,e} \in \mathbb{R}^{d^{\prime}\times |\mathcal{B}|}$ are I-int embedded representations for users, items, and bundles, respectively. $Y$ and $Z$ represent the observed user-item interactions and bundle-item affiliations. $E_{i}^{I-int} \in \mathbb{R}^{d \times |\mathcal{I}|}$ is the initial item embeddings. \(\mathcal{AGG}(\cdot)\) denotes an aggregation function, such as mean pooling. 

Analogous to the bundle-level interaction view, LightGCN is used to capture the user-item CF signals. The details of graph propagation in the item-level interaction view as follows:
\begin{equation}\begin{cases}\label{item_view_colaborative}
e_u^{I-int(k)}=\sum_{i\in\mathcal{N}_u}\frac{1}{\sqrt{|\mathcal{N}_u|}\sqrt{|\mathcal{N}_i|}}e_i^{I-int(k-1)}, \quad r_u^{I-int,e}=\textstyle \frac{1}{K}\sum_{k=0}^Ke_u^{I-int(k)} \\
e_i^{I-int(k)}=\sum_{u\in\mathcal{N}_i}\frac{1}{\sqrt{|\mathcal{N}_i|}\sqrt{|\mathcal{N}_u|}}e_u^{I-int(k-1)}, \quad r_i^{I-int,e}=\textstyle \frac{1}{K}\sum_{k=0}^Ke_i^{I-int(k)} & 
\end{cases}\end{equation}
where $e_{u}^{I-int(k)}$, $e_{i}^{I-int(k)} \in \mathbb{R}^{d}$ are the $k$-th layer’s embeddings for user $u$ and item $i$; $\mathcal{N}_u$ and $\mathcal{N}_i$ are the neighbors of the user $u$ and item $i$ in the user-item graph. We pool different layers of embeddings into I-int embedded representations $r_u^{I-int,e}$ and $r_i^{I-int,e}$. Then, we aggregate the I-int items embedded representations into I-int bundle embedded representations by mean pooling: 
\begin{equation}\label{agg}
r_{b}^{I-int,e}=\frac{1}{|\mathcal{C}_b|}\textstyle \sum_{i\in \mathcal{C}_b}r_{i}^{I-int,e}
\end{equation}
where $r_{b}^{I-int,e}$ is the I-int bundle embedded representation and $\mathcal{C}_b$ is the item set of bundle $b$.

\section{The MoDiffE Framework}\label{sec:framework}

\begin{figure}[t]
    \includegraphics[width=\textwidth]{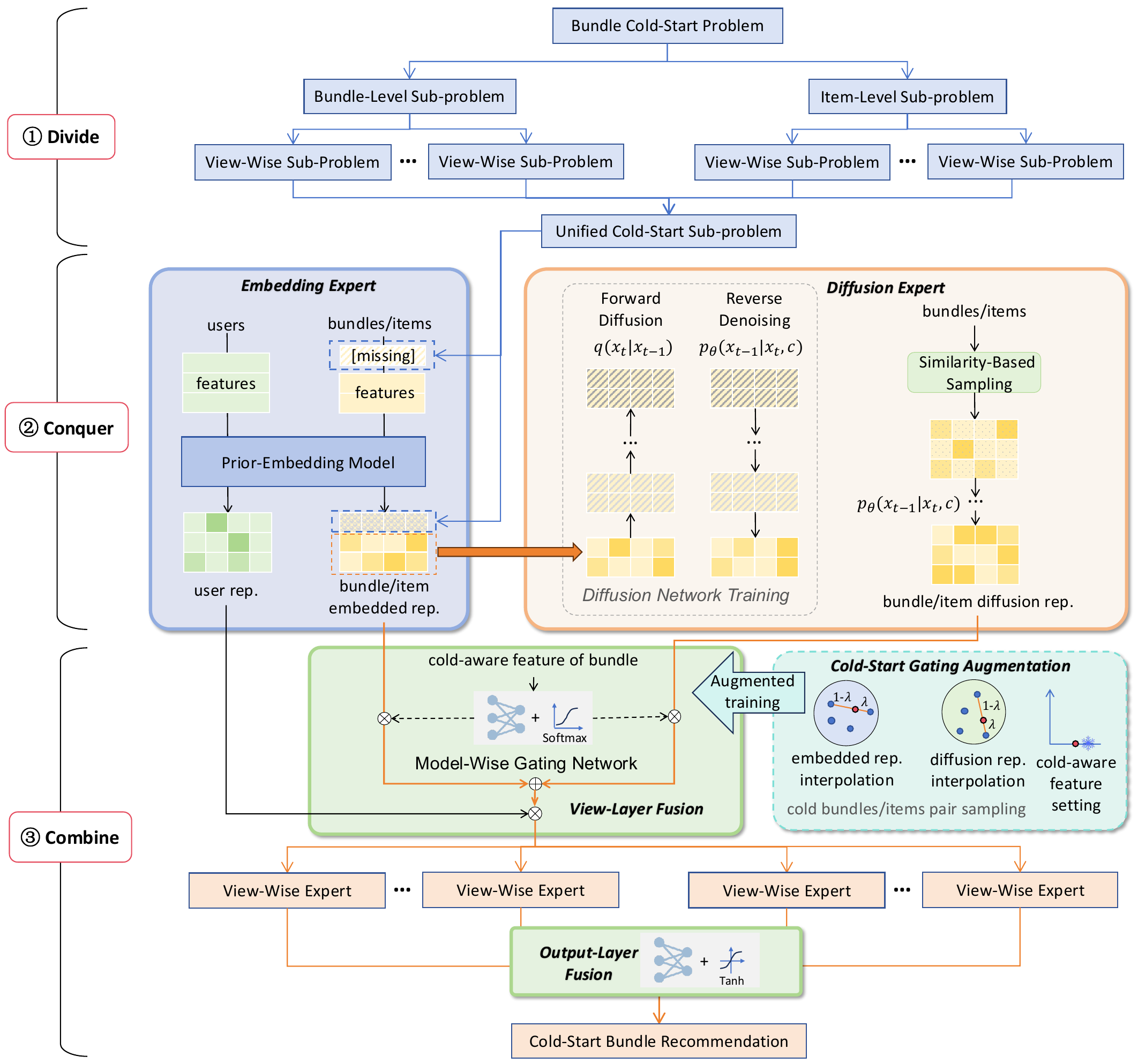}
    \caption{The MoDiffE framework}
    \label{framework}
\end{figure}

In this section, we will detail the design of our Mixture of Diffusion Experts (MoDiffE) framework illustrated in Figure~\ref{framework}. 
MoDiffE employs a divide-and-conquer strategy and follows three steps: (1) divide bundle cold-start problem into independent sub-problems and unify them into the same form (Section~\ref{sec:divide}); (2) conquer all sub-problems uniformly based on a diffusion-based method (Section~\ref{sec:conquer}); (3) combine the results of all sub-problems into the solution of the original cold-start problem by a cold-aware hierarchical MoE (Section~\ref{sec:combine}). Moreover, MoDiffE adopts a multi-stage decoupled pipeline for training and uses a cold-start gating augmentation method to enable the training of gating on cold bundles (Section~\ref{sec:training}). 
Next, we will explain them one by one. 

\subsection{Divide and Unify Sub-problems}\label{sec:divide}

\subsubsection{Divide}
Prior-embedding models fundamentally provide embedded representations of bundles but suffer from the cold-start problem. 
The dual-level multi-view complexity makes the bundle cold-start problem difficult to solve directly. 
Motivated by the divide-and-conquer strategy, MoDiffE divides the problem into smaller, more manageable sub-problems. 
The bundle representation learning has the dual-level multi-view characteristic, where each layer (i.e., bundle and item) and view (e.g., interaction view and content view) may encounter distinct cold-start problems. 
This division follows the order of level first and then view to obtain multiple sub-problems. 
By dividing into these independent sub-problems, we can address the unique problems associated with each level and view without being overwhelmed by the overall complexity. 
This approach simplifies the dual-level multi-view complexity of the bundle cold-start problem and reduces the overall complexity of finding a solution, making it more efficient and targeted. 

\subsubsection{Unify}
Since the number of sub-problems grows multiplicatively with the combination of levels and views, individually addressing each cold-start sub-problem may introduce excessive complexity and inefficiency. 
Therefore, it is essential to find a more efficient and unified approach to tackle this problem. 
Through analysis of advanced bundle recommendation models, we observe that they mainly follow a prior-embedding paradigm: 
a representation learning model is trained to learn a vectorized representation of an entity given the specific features. 
This paradigm is expressed in a specific view as follows: 
\begin{equation}
r = f_{\theta}^{v}(h)
\end{equation}
where $f_{\theta}^{v}$ is the prior-embedding model in view $v$, $h$ is the feature of bundle/item, and $r$ is the vectorized representation of bundle/item. 
The item level has one more aggregation operation than the bundle level, but the root cause of its cold start is still in representation learning. 
Based on the above observation, the cold-start sub-problem across different levels and views can be summarized as a unified sub-problem that the poor representation of feature-missing bundles/items in the prior-embedding model. This can be formulated as:
\begin{equation}\label{eq:unified_subproblem}
r = f_{\theta }^{v}(h=\varnothing )=\varnothing
\end{equation}
This problem is shown in the prior-embedding model part of the Figure~\ref{framework}.

\subsection{Conquer by Diffusion Model}\label{sec:conquer}
The bundle cold-start problem is divided into multiple unified sub-problems as Eq~\ref{eq:unified_subproblem}. Solving this unified sub-problem can solve all cold-start sub-problems across different levels and views. 
In this section, we propose a diffusion-based unified method for all cold-start sub-problems. 

\subsubsection{Diffusion Model}
Diffusion models have emerged as the state-of-the-art of generative modeling paradigm, achieving substantial success in diverse fields, including computer vision~\cite{diffusion_cv_2021a,diffusion_cv_2021b,diffusion_cv_DDPM_2020} and natural language processing~\cite{diffusion_nlp_2022,diffusion_nlp_2023} and other areas~\cite{diffusion_audio_2020}. 
Unlike other earlier generative models such as VAEs~\cite{VAEs} and GANs~\cite{GAN},  diffusion models leverage a denoising framework that effectively reverses a multi-step noising process to generate synthetic data that aligns closely with the distribution of the training data. 
They have demonstrated superior performance in high-quality data generation compared with VAEs and GAN models, owing to their ability to achieve more stable training and diversity in generation~\cite{diffusion_cv_DDPM_2020}. 
Denoising diffusion probabilistic model (DDPM)~\cite{diffusion_cv_DDPM_2020} is the most commonly used diffusion model framework that consists of two parts: (1) the forward diffusion process that gradually transforms the data into pure noise with Gaussian noise, and (2) the reverse denoising process that aims to recover the original data via deep neural networks. 
The training of the diffusion model is based on the diffusion-denoise processes, while only the reverse process is required for inference. 
Then, we will introduce these two processes, training and inference of our method in detail, taking the bundle-level interaction view as an example.

\subsubsection{Forward Diffusion Process}
Let $x_{0} \sim q(x_0)$ denotes the original data samples. The forward diffusion process is a Markov Chain that gradually adds Gaussian noise to $x_0$, ultimately aiming to progress towards convergence with the standard Gaussian distribution (i.e., pure  noise). 
It can be formulated as follows:
\begin{equation}\begin{aligned}
 & q(x_{1:T}|x_0):=\textstyle \prod_{t=1}^Tq(x_t|x_{t-1}) \\
 & q(x_t|x_{t-1}):=\mathcal{N}(x_t;\sqrt{1-\beta_t}x_{t-1},\beta_t\mathbf{I})
\end{aligned}\end{equation}
where $T$ is the number of steps to add noise and $\beta_t \in (0, 1)$ is the variance schedule that controls the noise scales added at step $t$. If $T\longrightarrow \infty $, $x_T$ approaches a standard Gaussian distribution~\cite{diffusion_cv_DDPM_2020}. $\mathbf{I}$ is the identity matrix with the same dimension as the input data $x_{t-1}$, and $\mathcal{N}(x; \mu, \sigma \mathbf{I})$ is a Gaussian distribution of $x$ with the mean $\mu$ and the standard deviation $\sigma \mathbf{I}$.
The key observation is that we can represent any $x_t$ using the original sample $x_0$ with the reparameterization trick~\cite{VAEs}. The forward diffusion process is transformed as follows: 
\begin{equation}\label{add_noise}
q(x_t|x_0)=\mathcal{N}(x_t;\sqrt{\bar{\alpha}_tx_t},(1-\bar{\alpha}_t)\mathbf{I})
\end{equation}
where $\alpha_t = 1 - \beta_t$ and $\bar{\alpha}_t=\textstyle \prod_{s=1}^t\alpha_s$. Hence, we can re-parameterize $x_t$ as following:
\begin{equation}x_t=\sqrt{\bar{\alpha}_t}x_0+\sqrt{1-\bar{\alpha}_t}\epsilon\end{equation}
where $\epsilon\sim\mathcal{N}(0,\mathbf{I})$ is the Gaussian noise. To regulate the added noises in $x_{1:T}$, there are some existing noise schedule for $1-\bar{\alpha}_T$ to choose from, such as linear noise schedule, cosine linear noise schedule, exp noise schedule, etc. 

In the bundle-level interaction view, give a warm bundle $b \in \mathcal{B}_{w}^{B\text{-}int}$ and its B-int embedded representation $r_{b}^{B-int,e}$. We set the initial state $x_0 = r_{b}^{B-int,e}$ and use Eq.~\ref{add_noise} to corrupt it.

\subsubsection{Reverse Denoising Process}
The reverse process aims to recover the original data sample $x_0$ from the completely perturbed $x_T\sim\mathcal{N}(0,\mathbf{I})$ through a series of Markov chain based transformations.  
Unfortunately, estimating the reverse denoising process $p(x_{t-1} | x_t )$ is quite difficult because it requires modeling the entire dataset's distribution. Therefore, a learnable neural network is applied to estimate the corresponding conditional distribution $p_\theta(x_{t-1}|x_t)$: 
\begin{equation}\label{reverse_denoising}
p_\theta\left(x_{t-1}\mid x_t\right)=\mathcal{N}\left(x_{t-1};\boldsymbol{\mu}_\theta\left(x_t,t\right),\sigma_\theta\left(x_t,t\right)\mathbf{I}\right)
\end{equation}
where $\mu_\theta(x_t, t)$ and $\sigma_\theta(x_t, t)$ are the mean and covariance of the Gaussian distribution predicted by a neural network with parameters $\theta$. The goal is to learn and approximate the data distribution via the model with parameters $\theta$ during the reverse denoising process. 

Since each bundle in the recommendation system is a unique entity, its representation should emphasize determinism rather than diversity as in the image generation field. 
The vanilla diffusion model generates the data samples based on the learned distribution of the source data without any explicit guidance or conditions.
Fortunately, the diffusion models can generate the data samples not only from an unconditional distribution, but also from a conditional distribution given a condition~\cite{diffusion_rec_survey_2025,diffusion_survey_2023}. 
To generate a deterministic representation of a particular bundle, we take the following conditional denoising transition step: 
\begin{equation}\label{conditional_reverse}
p_\theta\left(x_{t-1}\mid x_t, c\right)=\mathcal{N}\left(x_{t-1};\boldsymbol{\mu}_\theta\left(x_t,c,t\right),\sigma_\theta\left(x_t,c,t\right)\mathbf{I}\right)
\end{equation}
where $c$ is the condition of bundle $b$ to guide the reverse denoising process.
Condition, typically a feature representation, is not restricted to current view features and can incorporate labels, text, relationships, or other features~\cite{diffusion_survey_2023,diffusion_rec_survey_2024}. 
This paper treats the items within a bundle as its features and conduct pre-training based on bundle-item affiliations to get these feature representations. 

\begin{algorithm}[t]
\caption{Diffusion Model Training for View $v$}
\label{alg:diffusion_training}
\KwIn{Warm bundle set $\mathcal{B}^{v}_{w}$, noise steps $T$, diffusion model $f_\theta$}
\KwOut{Trained diffusion model parameters $\theta$}

  \ForEach{$b \in \mathcal{B}^{v}_{w}$}{
    $x_0 = r_b^{v,e}$\;
    Sample $t \sim \mathrm{Uniform}([1,T])$\;
    Add noise: $x_t = \sqrt{\bar\alpha_t}\,x_0 + \sqrt{1-\bar\alpha_t}\,\epsilon$, $\epsilon\sim\mathcal{N}(0,I)$\;
    Predict $\hat x_0 = f_\theta(x_t, c, t)$\;
    Compute step loss $L_t = \|x_0 - \hat x_0\|^2$\;
    Accumulate $\mathcal{L}^{\mathrm{diff}} \mathrel{+}= L_t$\;
  }
  Update $\theta$ by minimizing $\mathcal{L}^{\mathrm{diff}}$\;

\end{algorithm}

\subsubsection{Training}\label{sec:diffusion_training}
The objective of diffusion model training is to capture the underlying distribution of warm representations of warm bundles/items, enabling it to generate representations close to the actual distribution.
This is achieved by optimizing the negative log-likelihood through a variational lower bound. It can be further expressed as a combination of KL-divergence and entropy terms~\cite{diffusion_cv_DDPM_2020}. Here, we focus on the simplification~\cite{diffusion_survey_2023,diffusion_rec_survey_2024,diffusion_rec_survey_2025} and this loss $L_t$ at step $t$ of reverse denoising process can be written in the following form:
\begin{equation}\label{loss:diff}
L_{t} =\mathbb{E}_{t\sim[1,T],x_0,\epsilon_t}\left[\|x_0-\hat{x}_0\|^2\right] \\
\end{equation}
where $x_0$ is the original representation and $\hat{x}_0 = f_\theta\left(x_t,c, t\right) = MLP\left(x_t,c, t\right)$ is the generated representation. 
The diffusion model training for view $v$ as shown in Algorithm~\ref{alg:diffusion_training} (we take bundle as an example in Algorithm~\ref{alg:diffusion_training} and~\ref{alg:diffusion_inference}. The training and inference of item are similar.). 
Instead of following the vanilla diffusion model to predict the added noise for training, we directly predict the original representation. 
This is because applying the L2 loss directly to the representation helps to stabilise the scale of representation, making the generated representations more comparable.

\subsubsection{Inference}\label{inference}
The process of generating diffusion representation is shown in Algorithm~\ref{alg:diffusion_inference}. 
During inference, the diffusion model transforms anchor representations into deterministic representations via the reverse denoising process. 
Unlike the vanilla diffusion model, which samples Gaussian noise as the initial representation for denoising, the diffusion model in MoDiffE adopts a similarity-based sampling method as Eq.~\ref{similarity_sampling}.  
Specifically, give the input bundle $b$, we calculate similarities with the warm bundles based on item compositions, then select the top-n bundles and calculate the mean of their embedded representations as the anchor representation $r_{b}^{\text{anchor}}$. 
The similarity-based sampling method allows the generation process to have a better starting point closer to the actual representation, thus improving the stability of the generation process.
\begin{equation}\label{similarity_sampling}
r_{b}^{\text{anchor}} = MeanPooling(TopN(Similarity(b, \mathcal{B}^{B\text{-}int}_w \setminus b,  Z), R_b^{B-int,e})
\end{equation}

With the anchor representation $r_{b}^{\text{anchor}}$ of input bundle, the diffusion model applies a $T'$-step reverse denoising process (Eq.~\ref{conditional_reverse}) to gradually construct the representation step by step as follows:

\begin{equation}
r_{b}^{\text{anchor}} = x_{T'} \longrightarrow x_{T'-1} \longrightarrow \cdots \longrightarrow x_1 \longrightarrow x_0 = r_b^{B-int,d} \end{equation}
where $r_b^{B-int,d}$ is the bundle-level interaction-view diffusion representation of bundle $b$. 

In the item-level interaction view, it first generates I-int diffusion representations of items $R^{I-int,d}_i$ same to process of bundle, and then aggregates them into I-int diffusion representation $r_b^{I-int,d}$ for bundle $b$ according to Eq.~\ref{agg}. The similarity-based sampling method of item and item condition is similar to the bundle layer.

The inference can be time-consuming. We adopt a fast inference solver from existing work~\cite{DPM-Solver} for diffusion ODEs to speed up the sampling process with much fewer steps $T'$.

\begin{algorithm}[t]
\caption{Inference for View $v$}
\label{alg:diffusion_inference}
\KwIn{Input bundle $b$, trained diffusion model $f_\theta$, warm bundle set $\mathcal{B}_{w}^{v}$, inference steps $T'$}
\KwOut{Diffusion representation $r_b^{v,d}$}

Compute anchor $r_{b}^{\text{anchor}} = MeanPooling(TopN(Similarity(b, \mathcal{B}^{v}_w \setminus b, Z), R_b^{v,e})$\;

Set $x_{T'} \leftarrow r_{b}^{anchor}$\;

\For{$t = T'$ \KwTo $1$}{
  Predict $x_{t-1} = f_\theta(x_t, c, t)$\;
}

\Return{$r_b^{v,d} = x_0$}
\end{algorithm}

\subsection{Combine by Cold-Aware Hierarchical MoE}\label{sec:combine}
The MoE is utilized to combine results of all sub-problems into final recommendations, and prior-embedding and diffusion models of each view serve as experts. 
The original dense MoE model uses a common gating network to fuse results from all experts and can be formulated as follows:
\begin{equation}\label{original_MoE}
r_b=\sum_{i=1}^mg(b)_if_i(b)
\end{equation}
where $\sum_{i=1}^mg(b)_i=1$ and $g(b)_i$, the $i$-th logit of the output of $g(b)$, indicates the weight for expert $f_i$. The final representation of bundle $b$, denoted as $r_{b}$, is a weighted sum of the outputs of all $m$ experts. However, there are two problems in directly applying this gating method to MoDiffE.

On the one hand, in the context of bundle recommendation tasks, the objective is to estimate the likelihood of user interaction with each candidate bundle and subsequently rank them to produce the final bundle recommendation list. 
This task requires comparative predictions of all candidate bundles, rather than independent predictions for each bundle. Therefore, the representations of all bundles need to follow a consistent and stable distribution. 
However, MoDiffE incorporates heterogeneous experts at different views to learn different representations. 
It is difficult to achieve stability in the final representation distribution by directly fusing the outputs of all experts from divers layers and views using the unified gating method described in Eq.~\ref{original_MoE}. 

On the other hand, the bundle cold-start problem occurs in dual-level multi-view representation learning, and different cold-start problems at different views lead to different requirements for representation learning. The unified gating method makes it difficult to handle bundles in diverse cold-start situations adaptively. 

To address the above two problems, we propose a cold-aware hierarchical MoE. It involves multiple gating networks that first perform view-layer gating for heterogeneous expert representations fusion, then proceed to output-layer gating to fuse predictions from different views.

\subsubsection{View-Layer Gating for Representation Fusion}
The gating networks of different views input view-specific cold-aware features for more effective adaptive processing with various cold-start situations. 
The view-layer gating can be formulated as: 
\begin{equation}\label{hierarchical_fusion}
\begin{aligned}
    r_b^{v} = \sum_{i=1}^{n} g^{v}(a^{v}_b)_i f^{v}_i(b),
\end{aligned}
\end{equation}
where $g^{v}(\text{a}^{v}_b)_i$ represents the weight assigned to the $i$-th expert $f^{v}_i$ by the gating network at view $v$.
$a^{v}_b$ is the cold-aware feature of input bundle $b$ in view $v$. 
In this paper, cold-aware features are the number of user interactions of bundles and items. 

The gating networks are simply linear transformations of the input with a normalization function. Since $g^{v}$ is used to fuse representations from different experts, which are in a competitive relationship, we employ the $Softmax$ function to normalize the weights accordingly. $g^{v}$ is formulated as Eq.~\ref{gating} where $W^{v}$ is a trainable matrix. 
\begin{equation}\label{gating}
g^{v}(a_b^{v})=\mathrm{Softmax}(W^{v}a_b^{v})
\end{equation}

\subsubsection{Output-Layer Gating for Prediction Fusion}
We utilize the inner product to estimate the likelihood of interaction between user $u$ and bundle $b$ at different views, and then fuse them into final predictions by a gating network. It can be formulated as follows: 
\begin{equation}\label{prediction}
y_{u,b} = \sum_{v\in V} g^{out}(a_{b}^{out})_v {r_u^{v}}^{T} r_b^{v}
\end{equation}
where $y_{ub}$ is the final prediction and $g^{out}(a_{b}^{out})_v$ means the weight assigned to the prediction of view $v \in V$. $r_u^{v}$ is from Eq.~\ref{bundle_view_colaborative} and Eq.~\ref{item_view_colaborative}. $a_{b}^{out}$ is the feature of bundle $b$ for output-layer gating. 

Different views predict user interactions from different perspectives. 
They are not competitive but relatively cooperative. 
Existing bundle recommendation models~\cite{BGCN,CrossCBR,MIDGN,MultiCBR,HyperMBR,AMCBR} implicitly learn cooperative relationships between different views and ultimately reflect in the representation scales of different views. 
MoDiffE enhances the original bundle model in multiple views but disturbs the cooperative relationship between different views. 
Therefore, we set up the output-layer gating network with the $Tanh$ activation function to explicitly reestablish new cooperative relationships between different views. 
$g^{out}(a_b^{out})$ is formulated as Eq.~\ref{gate_final} where $W^{out}$ is a learnable matrix and $a_b^{out}$ is the feature of input bundle for output-layer fusion. We set $a_b^{out}$ to be the concatenation of the representations of a bundle of all views. 
\begin{equation}\label{gate_final}
g^{out}(a_b^{out})=\mathrm{Tanh}(W^{out}a_b^{out})
\end{equation}

\subsection{Framework Training}\label{sec:training}
\subsubsection{Multi-Stage Decoupled Pipeline Training}
The $L_2$ distance-based loss used in the diffusion model and the inner product-based loss commonly used in the recommendation system are inconsistent in optimization direction and difficult to optimize simultaneously~\cite{diffusion_rec_DimeRec_2025}.
To address this issue, our MoDiffE adopts a multi-stage decoupled pipeline training, as outlined in Algorithm~\ref{alg:multi_stage_training}. 
Multi-stage training enhances the stability of both diffusion models and gating networks training. Meanwhile, decoupling the modules allows for greater flexibility in module selection and improves the overall adaptability of the MoDiffE framework.

\begin{algorithm}[t]
\caption{Multi-Stage Decoupled Pipeline Training}
\label{alg:multi_stage_training}
\KwIn{Original training set $Q$, interpolation sampling ratio $\eta$}
\KwOut{Trained MoDiffE framework}

\textbf{Stage 1: Train Prior-Embedding Model} \\
Train a prior-embedding model for bundle recommendation on $Q$ using BPR loss Eq.~\ref{loss:BPR}; \\

\textbf{Stage 2: Train Diffusion Models for Each View} \\
Train diffusion models on warm embedded representations using $L_2$ loss Eq.~\ref{loss:diff};

\textbf{Stage 3: Train Cold-Aware Hierarchical MoE } \\
Compute number of augmentation set $N = \eta \cdot |Q|$ and sample bundles for interpolation;\\
Interpolate pseudo cold bundles by Eq.~\ref{interpolation};

Train gating networks using augmented training set $Q \cup S$ with BPR loss as Eq.~\ref{loss:moe};

\Return{Trained MoDiffE}
\end{algorithm}

Specifically, a fundamental bundle recommendation model (prior-embedding model) is trained by the Bayesian Personalized Ranking (BPR) loss~\cite{BPR} based on the original training set as follows:. 
\begin{equation}\label{loss:BPR}
\mathcal{L}^{BPR}=\sum_{(u,b,b^{\prime})\in Q}-\mathrm{ln}\sigma(y_{u,b}-y_{u,b^{\prime}})
\end{equation}
where $Q=\{(u,b,b^{\prime})|u\in\mathcal{U},b,b^{\prime}\in\mathcal{B},x_{ub}=1,x_{ub^{\prime}}=0\}$, $\sigma(\cdot)$ is the Sigmoid function. 

Then, diffusion models for each views are trained as described in Section~\ref{sec:diffusion_training}.

Next, we train gating networks in cold-aware hierarchical MoE by recommendation loss (i.e. BPR loss). However, the cold bundle is not present in the training, gating networks in MoE are unable to handle it accurately during inference. 
Therefore, we propose a cold-start gating augmentation method (details in Section~\ref{sec:Cold-Start_Gating_Augmentation}) to synthesize pseudo cold bundles to expand the training set. The training loss of gating networks is as follows: 
\begin{equation}
\mathcal{L}^{BPR}=\sum_{(u,b,b^{\prime})\in Q \cup S}-\mathrm{ln}\sigma(y_{u,b}-y_{u,b^{\prime}})
\end{equation}\label{loss:moe}
where $S = \{(u, <b_x, b_y>, <b_x^{\prime}, b_y^{\prime}>)|u\in\mathcal{U}; b, b^{\prime}\in \mathcal{B}_{c}^{B\text{-}int} \}$. $\eta = |S|/|Q|$ is the interpolation sampling ratio to control the amount of interpolated data. 

\subsubsection{Cold-Start Gating Augmentation}\label{sec:Cold-Start_Gating_Augmentation}
If the cold bundle does not appear in the training set, gating networks in MoE are unable to handle it accurately during inference. 
To enhance gating for cold bundles, we propose a cold-start gating
augmentation method to interpolation-based synthesize pseudo cold bundles during training. 
Compared to augmenting the dataset by mining unobserved possible user-bundle interactions~\cite{diff_for_aug_2024}, interpolation-based constructing pseudo bundles does not introduce a negative impact on the distribution learned by the original dataset. 
Because the former is trained based on real IDs and inevitably introduces noise by augmentation, while the latter is trained based on pseudo IDs and it will not affect the representation of the original bundles. 

Specifically, given a bundle pair $<b_x,b_y>$ sampled from the cold bundle set $\mathcal{B}_{c}^{B\text{-}int}$, we use a linear interpolation method~\cite{mixup_2018,interpolation_ood_2023} to construct the representations and cold-aware feature of the pseudo cold bundle $b_{xy}$. The bundle-level interpolation can be formulated as follows:
\begin{equation}\label{interpolation}
\begin{cases}
\tilde{r}^{v,e}_{b_{xy}}\triangleq\lambda\cdot r^{v,e}_{b_x}+(1-\lambda)\cdot r^{v,e}_{b_y} \\
\tilde{r}^{v,d}_{b_{xy}}\triangleq\lambda\cdot r^{v,d}_{b_x}+(1-\lambda)\cdot r^{v,d}_{b_y} \\
\tilde{a}_{b_{xy}}^{v} = 0 & 
\end{cases}\end{equation}
where the linear interpolation ratio $\lambda \in [0,1]$ is drawn from a Beta distribution Beta$(\alpha,\alpha)$. $\tilde{r}^{v,e}_{b_{xy}}$ and $\tilde{r}^{v,d}_{b_{xy}}$ refer to the interpolated embedded and diffusion representations at view $v$. 
$\tilde{a}_{b_{xy}}^{v}$ is the cold-aware feature at view $v$, which is set to 0 to simulate the cold-start case. 
The amount of interpolated data is controlled by the interpolation sampling ratio $\eta$, which represents the ratio of the amount of interpolated data to the amount of original data. Note that the item-layer representation is also interpolated to maintain consistency across layers.

\section{Experiments}
\subsection{Experimental Setup}

\subsubsection{Dataset}
Existing research on bundle recommendation and cold-start problem mainly focuses on interaction data. For the comparability of experiments, we follow this setting and conduct our experiments based on the interaction view data.
We follow the three public real-world bundle datasets from the existing cold-start bundle recommendation research~\cite{bundle_cold_start_CoHeat_2024}, i.e., Youshu~\cite{DAM}, NetEase~\cite{bundle_rec_2017b}, and iFashion~\cite{CrossCBR}, corresponding to domains of book list, music playlist and fashion outfit, respectively. 
Their statistics are shown in Table~\ref{tab:dataset}. 

\begin{table}[h]
\centering
\footnotesize
\caption{Statistics of three real-world datasets in our experiments.}
\label{tab:dataset}
\begin{tabular}{l|rrr|rrr|c}
\hline
\textbf{Dataset} & \textbf{User} & \textbf{Bundle} & \textbf{Item} & \textbf{U-B Interaction} & \textbf{U-I Interaction} & \textbf{B-I Affiliation} & \textbf{Avg. Bundle Size} \\
\hline
Youshu & 8,039 & 4,771 & 32,770 & 51,377 & 138,515 & 176,667 & 37.03 \\
NetEase & 18,528 & 22,864 & 123,628 & 302,303 & 1,128,065 & 1,778,838 & 77.80 \\
iFashion & 53,897 & 27,694 & 42,563 & 1,679,708 & 2,290,645 & 106,916 & 3.86 \\
\hline
\end{tabular}
\end{table}

\subsubsection{Experimental Protocol}
We follow the experimental protocol from the existing cold-start bundle recommendation research CoHeat~\cite{bundle_cold_start_CoHeat_2024}. 
All-unrated-item~\cite{evaluation_protocol} as a widely used recommendation evaluation protocol is adopted in this paper: for each user, we retain all candidate bundles with whom she does not interact within the training set. Top-K bundle recommendation is regarded as the specific task of this paper: recommend a ranked list of $K$ bundles from the candidate bundles to a user. We follow existing cold-start bundle recommendation research~\cite{bundle_cold_start_CoHeat_2024} to set $K$ to 20. 

Two widely used metrics~\cite{BGCN,CrossCBR,HyperMBR,MIDGN,MultiCBR,bundle_cold_start_CoHeat_2024,AMCBR}, Recall@K and NDCG@K, are employed to evaluate the top-K recommendation performance related to personalization. Recall measures the ratio of test bundles within the top-K ranking list, and NDCG (Normalized Discounted Cumulative Gain) accounts for the position of the hits by assigning higher scores to those at the top ranks. 

We conducted experiments across cold-start, all-bundle, and warm-start scenarios, consistent with previous studies~\cite{bundle_cold_start_CoHeat_2024,item_cold_start_CLCRec_2021}. In the warm-start scenario, user-bundle interactions are divided into a 7:1:2 ratio for training, validation, and testing. For the cold-start scenario, bundles are split in a 7:1:2 ratio, with corresponding user interactions allocated to training, validation, and testing. 
The all-bundle scenario is a blend of warm-start and cold-start scenarios. Interactions are split in 7:1:2 ratio with a half for warm-start bundles and the other half for cold-start bundles for the validation and testing set. 
We use the dataset~\footnote{https://github.com/snudatalab/CoHeat} released by CoHeat to keep the splitting consistent. 

\subsubsection{Cold-Start Situation Analysis}

\begin{table}[t]
\caption{Cold-start situation analysis of three datasets at different scenario settings based on interaction views, where the first bracket (\textcolor{violet}{violet}) indicates the ratio of bundles/items, and the second bracket (\textcolor{cyan}{blue}) indicates the ratio of bundle-related interactions in the test set. BL and IL means bundle-level and item-level.}
\footnotesize
\label{tab:data_analysis}
\renewcommand{\arraystretch}{1.12}
\begin{tabular}{cc|lll}
\hline \hline
\multicolumn{2}{c|}{\textbf{Dataset}}                                                     & \multicolumn{1}{l}{\textbf{Youshu}}              & \multicolumn{1}{l}{\textbf{NetEase}}             & \multicolumn{1}{l}{\textbf{iFashion}}            \\ \hline
\multicolumn{2}{c|}{Item}                                                                 & 32,770                                            & 123,628                                           & 42,563                                            \\
\multicolumn{2}{c|}{Warm Item $\mathcal{I}_{w}$}                                                      & 21,034 \textcolor{violet}{(64.19\%)}                                  & 90,520 \textcolor{violet}{(73.22\%)}                                  & 40,378 \textcolor{violet}{(94.87\%)}                                  \\
\multicolumn{2}{c|}{Cold Item $\mathcal{I}_{c}$}                                                      & 11,736 \textcolor{violet}{(35.81\%)}                                  & 33,108 \textcolor{violet}{(26.78\%)}                                  & 2,185 \textcolor{violet}{(5.13\%)}                                    \\ \hline
\multicolumn{2}{c|}{Bundle}                                                               & 4,771                                             & 22,864                                            & 27,694                                            \\
\multicolumn{2}{c|}{IL Warm Bundle $\mathcal{B}_{w}^{I-int}$}                                                     & 2,208 \textcolor{violet}{(46.28\%)}                                   & 4,596 \textcolor{violet}{(20.10\%)}                                   & 25,506 \textcolor{violet}{(92.10\%)}                                  \\
\multicolumn{2}{c|}{IL Cold Bundle $\mathcal{B}_{c}^{I-int}$}                                                     & 2,563 \textcolor{violet}{(53.72\%)}                                   & 18,268 \textcolor{violet}{(79.90\%)}                                  & 2,188 \textcolor{violet}{(7.90\%)}                                    \\ \hline \hline
\multicolumn{2}{c|}{\textbf{Scenario}}                                                    & \multicolumn{1}{l}{\textbf{Cold-Start Scenario}} & \multicolumn{1}{l}{\textbf{All-Bundle Scenario}} & \multicolumn{1}{l}{\textbf{Warm-Start Scenario}} \\ \hline 
\multicolumn{1}{c|}{\multirow{6}{*}{\textbf{Youshu}}}   & BL Warm Bundle $\mathcal{B}_{w}^{B\text{-}int}$              & 3,310 \textcolor{violet}{(0\%)} \textcolor{cyan}{(0\%)}                             & 3,768 \textcolor{violet}{(67.07\%)} \textcolor{cyan}{(43.92\%)}                           & 4,125 \textcolor{violet}{(82.79\%)} \textcolor{cyan}{(95.19\%)}                           \\
\multicolumn{1}{c|}{}                                   & BL Cold Bundle $\mathcal{B}_{c}^{B\text{-}int}$              & 1,461 \textcolor{violet}{(100\%)} \textcolor{cyan}{(100\%)}                         & 1,003 \textcolor{violet}{(32.93\%)} \textcolor{cyan}{(56.08\%)}                           & 646 \textcolor{violet}{(17.21\%)} \textcolor{cyan}{(4.81\%)}                             \\ \cline{2-5} 
\multicolumn{1}{c|}{}                                  & $\mathcal{B}_{w}^{B\text{-}int} \cap \mathcal{B}_{w}^{I-int}$ & 1,530 \textcolor{violet}{(0\%)} \textcolor{cyan}{(0\%)}                             & 1,724 \textcolor{violet}{(24.82\%)} \textcolor{cyan}{(13.40\%)}                           & 1,864 \textcolor{violet}{(31.87\%)} \textcolor{cyan}{(30.27\%)}                           \\
\multicolumn{1}{c|}{}                                   & $\mathcal{B}_{w}^{B\text{-}int} \cap \mathcal{B}_{c}^{I-int}$ & 1,780 \textcolor{violet}{(0\%)} \textcolor{cyan}{(0\%)}                             & 2,044 \textcolor{violet}{(42.24\%)} \textcolor{cyan}{(30.52\%)}                           & 2,261 \textcolor{violet}{(50.92\%)} \textcolor{cyan}{(64.92\%)}                           \\
\multicolumn{1}{c|}{}                                   & $\mathcal{B}_{c}^{B\text{-}int} \cap \mathcal{B}_{w}^{I-int}$ & 678 \textcolor{violet}{(45.08\%)} \textcolor{cyan}{(32.40\%)}                            & 484 \textcolor{violet}{(15.27\%)} \textcolor{cyan}{(17.68\%)}                            & 344 \textcolor{violet}{(8.65\%)} \textcolor{cyan}{(2.43\%)}                              \\
\multicolumn{1}{c|}{}                                   & $\mathcal{B}_{c}^{B\text{-}int} \cap \mathcal{B}_{c}^{I-int}$ & 783 \textcolor{violet}{(54.92\%)} \textcolor{cyan}{(67.60\%)}                            & 519 \textcolor{violet}{(17.67\%)} \textcolor{cyan}{(38.40\%)}                            & 302 \textcolor{violet}{(8.56\%)} \textcolor{cyan}{(2.38\%)}                              \\ \hline
\multicolumn{1}{c|}{\multirow{6}{*}{\textbf{NetEase}}}  & BL Warm Bundle $\mathcal{B}_{w}^{B\text{-}int}$              & 14,971 \textcolor{violet}{(0\%)} \textcolor{cyan}{(0\%)}                            & 17,716 \textcolor{violet}{(80.26\%)} \textcolor{cyan}{(49.66\%)}                          & 20,395 \textcolor{violet}{(95.25\%)} \textcolor{cyan}{(98.52\%)}                          \\
\multicolumn{1}{c|}{}                                   & BL Cold Bundle $\mathcal{B}_{c}^{B\text{-}int}$               & 7,893 \textcolor{violet}{(100\%)} \textcolor{cyan}{(100\%)}                         & 5,148 \textcolor{violet}{(19.74\%)} \textcolor{cyan}{(50.34\%)}                           & 2,469 \textcolor{violet}{(4.75\%)} \textcolor{cyan}{(1.48\%)}                             \\ \cline{2-5} 
\multicolumn{1}{c|}{}                                   & $\mathcal{B}_{w}^{B\text{-}int} \cap \mathcal{B}_{w}^{I-int}$ & 2,990 \textcolor{violet}{(0\%)} \textcolor{cyan}{(0\%)}                             & 3,531 \textcolor{violet}{(17.58\%)} \textcolor{cyan}{(13.48\%)}                           & 4,107 \textcolor{violet}{(20.48\%)} \textcolor{cyan}{(27.09\%)}                           \\
\multicolumn{1}{c|}{}                                   & $\mathcal{B}_{w}^{B\text{-}int} \cap \mathcal{B}_{c}^{I-int}$ & 11,981 \textcolor{violet}{(0\%)} \textcolor{cyan}{(0\%)}                            & 14,185 \textcolor{violet}{(62.68\%)} \textcolor{cyan}{(36.18\%)}                          & 16,288 \textcolor{violet}{(74.77\%)} \textcolor{cyan}{(71.43\%)}                          \\
\multicolumn{1}{c|}{}                                   & $\mathcal{B}_{c}^{B\text{-}int} \cap \mathcal{B}_{w}^{I-int}$ & 1,606 \textcolor{violet}{(19.73\%)} \textcolor{cyan}{(26.99\%)}                           & 1,065 \textcolor{violet}{(3.86\%)} \textcolor{cyan}{(12.52\%)}                            & 489 \textcolor{violet}{(0.75\%)} \textcolor{cyan}{(0.23\%)}                              \\
\multicolumn{1}{c|}{}                                   & $\mathcal{B}_{c}^{B\text{-}int} \cap \mathcal{B}_{c}^{I-int}$ & 6,287 \textcolor{violet}{(80.27\%)} \textcolor{cyan}{(73.01\%)}                           & 4,083 \textcolor{violet}{(15.87\%)} \textcolor{cyan}{(37.82\%)}                           & 1,980 \textcolor{violet}{(4.00\%)} \textcolor{cyan}{(1.25\%)}                             \\ \hline
\multicolumn{1}{c|}{\multirow{6}{*}{\textbf{iFashion}}} & BL Warm Bundle $\mathcal{B}_{w}^{B\text{-}int}$               & 19,385 \textcolor{violet}{(0\%)} \textcolor{cyan}{(0\%)}                            & 23,539 \textcolor{violet}{(88.58\%)} \textcolor{cyan}{(50.69\%)}                          & 27,694 \textcolor{violet}{(100\%)} \textcolor{cyan}{(100\%)}                        \\
\multicolumn{1}{c|}{}                                   & BL Cold Bundle $\mathcal{B}_{c}^{B\text{-}int}$               & 8,309 \textcolor{violet}{(100\%)} \textcolor{cyan}{(100\%)}                         & 4,155 \textcolor{violet}{(11.42\%)} \textcolor{cyan}{(49.31\%)}                           & 0 \textcolor{violet}{(0\%)} \textcolor{cyan}{(0\%)}                                \\ \cline{2-5} 
\multicolumn{1}{c|}{}                                   & $\mathcal{B}_{w}^{B\text{-}int} \cap \mathcal{B}_{w}^{I-int}$ & 17,845 \textcolor{violet}{(0\%)} \textcolor{cyan}{(0\%)}                            & 21,675 \textcolor{violet}{(82.05\%)} \textcolor{cyan}{(49.26\%)}                          & 25,506 \textcolor{violet}{(92.24\%)} \textcolor{cyan}{(97.07\%)}                          \\
\multicolumn{1}{c|}{}                                   & $\mathcal{B}_{w}^{B\text{-}int} \cap \mathcal{B}_{c}^{I-int}$ & 1,540 \textcolor{violet}{(0\%)} \textcolor{cyan}{(0\%)}                             & 1,864 \textcolor{violet}{(6.53\%)} \textcolor{cyan}{(1.43\%)}                             & 2,188 \textcolor{violet}{(7.76\%)} \textcolor{cyan}{(2.93\%)}                             \\
\multicolumn{1}{c|}{}                                   & $\mathcal{B}_{c}^{B\text{-}int} \cap \mathcal{B}_{w}^{I-int}$ & 7,661 \textcolor{violet}{(92.44\%)} \textcolor{cyan}{(96.92\%)}                           & 3,831 \textcolor{violet}{(10.54\%)} \textcolor{cyan}{(47.64\%)}                           & 0 \textcolor{violet}{(0\%)} \textcolor{cyan}{(0\%)}                                \\
\multicolumn{1}{c|}{}                                   & $\mathcal{B}_{c}^{B\text{-}int} \cap \mathcal{B}_{c}^{I-int}$ & 648 \textcolor{violet}{(7.56\%)} \textcolor{cyan}{(3.08\%)}                              & 324 \textcolor{violet}{(0.88\%)} \textcolor{cyan}{(1.67\%)}                              & 0 \textcolor{violet}{(0\%)} \textcolor{cyan}{(0\%)}                                \\ \hline \hline
\end{tabular}
\end{table}

The cold-start situation analysis of three experiment datasets at different scenario settings based on interaction views are shown in Table~\ref{tab:data_analysis}. 
Note that the original dataset does not contain bundle-level cold bundles (i.e. bundles without user interaction). In the experiment, it only appears after a specific data splitting. Based on the table, we have the following observations: 

\begin{itemize}
    \item \textbf{Item-level cold bundles are common in real-world datasets.} The preprocessing of the three datasets all conducted at least 10-core sparsity filtering~\cite{bundle_rec_2017b}. 
    However, these datasets still have a considerable number of item-level cold bundles, accounting for 79.7\% in NetEase and 53.72\% in Youshu. 
    This reflecting that the cold-start problem is common at two levels in the bundle recommendation task. 
    \item \textbf{Diverse cold-start situations of bundles coexist.} Different recommendation scenarios with different data sparsity are simulated based on different data splitting. In these scenarios, bundle-level and item-level cold-start problems are not independent of each other but cross-existing. In the all-bundle scenario, which is closer to the real scenario, the four types of bundles set account for a considerable proportion. For example, in Youshu, they account for 24.82\%, 42.24\%, 15.27\%, and 17.67\%, respectively. 
    This reflects the fact that bundle cold-start situations are diverse in bundle recommendations.
\end{itemize}

\subsubsection{Baselines}
We compare MoDiffE with existing state-of-the-art cold-start bundle recommendation methods and cold-start item recommendation methods. All baselines are as follows: 
\begin{itemize}
    \item \textbf{DropoutNet}~\cite{item_cold_start_DropoutNet_2017} (\textit{NIPS 2017}), a robustness-based method that randomly selects user-item pairs and sets their preference inputs to zero, forcing the model to rely solely on content information to reconstruct the relevance scores. 
    \item \textbf{CB2CF}~\cite{item_cold_start_CB2CF_2019} (\textit{RecSys 2019}), a constraint-based method that establishes a multi-view mapping between content-based (CB) features and collaborative filtering (CF) embeddings through cosine similarity optimization, enabling effective recommendations without requiring user interaction data. 
    \item \textbf{Heater}~\cite{item_cold_start_Heater_2020a} (\textit{SIGIR 2020}), a constraint-based method that transforms auxiliary representations to intermediate representations, then further refines them to final CF representations. Meanwhile, a similarity constraint is used to minimize the difference between the intermediate representations and pretrained CF representations.
    \item \textbf{GAR}~\cite{item_cold_start_GAR_2022} (\textit{SIGIR 2022}), a generative method that employs an adversarial training approach for the generator (content mapper) and the recommender, enabling the generator produces cold-start item representations that are close to the distribution of warm-start representations learned from historical interactions. 
    \item \textbf{CVAR}~\cite{item_cold_start_CVAR_2022} (\textit{SIGIR 2022}), a generative method that introduces Conditional Variational Autoencoder to learn a distribution over auxiliary information and generate desirable ID embeddings using a conditional decoder. 
    \item \textbf{CLCRec}~\cite{item_cold_start_CLCRec_2021} (\textit{MM 2021}), a contrastive learning-based method that learns cold representations from based on information theory. It maximizes the mutual information between feature representations and collaborative representations of items through contrastive learning. 
    \item \textbf{CCFCRec}~\cite{item_cold_start_CCFCRec_2023} (\textit{WWW 2023}), a contrastive learning-based method that jointly trains content-based and co-occurrence embeddings through contrastive learning, enabling implicit knowledge transfer to rectify cold-start embeddings during inference. 
    \item \textbf{CoHeat}~\cite{bundle_cold_start_CoHeat_2024} (\textit{WWW 2024}), a bundle cold-start method that designs a curriculum strategy that progressively transitions the training focus from user-bundle to user-item interactions, enabling the model to infer user preferences for new bundles based on user-item interactions.
\end{itemize}

\subsubsection{Implementation and Hyperparameter Configuration}
For implementation configuration, the contained items are considered as the content information of the bundle~\cite{bundle_cold_start_CoHeat_2024}. 
Since the MoDiffE framework adopts a multi-stage training strategy and the modules are decoupled, its implementation can be flexible and adaptable. 
In this paper, we choose CrossCBR~\cite{CrossCBR} as our foundational bundle recommendation model (prior-embedding model) because it is a typical dual-level multi-view bundle recommendation model. Its code uses its public version. 
DDPM~\cite{diffusion_cv_DDPM_2020}, the most commonly used diffusion model framework, is used to implement diffusion models at dual levels. 

For hyperparameter configuration, the embedding size is set to 64 to maintain consistency with baseline models. 
Hyperparameters of prior-embedding models follow released implementation of CrossCBR. 
The number of steps for the forward diffusion and reverse denoising processes is configured at $T=$ 500 and $T'=$ 20, respectively.
Noise schedule are tuned in linear noise schedule, cosine linear noise schedule, exp noise schedule. 
Learning rate and L2 regularization weight of diffusion models and gating networks are searched in the range of \{0.01, 0.003, 0.0001, 0.0003, 0.0001\} and \{1e-4, 1e-5, 1e-6, 1e-7, 0\}. 
The similarity-based sampling method is based on cosine similarity and uniformly selects the top $n$=5 bundles. 
$\alpha$ of Beta$(\alpha,\alpha)$ is set to 0.9. 
The interpolation sampling ratio $\eta$ is set to 0, 0.3, and 0.5 for warm-start, all-bundle, and cold-start scenarios.

\subsection{Main Result}

\begin{table}[t]
\caption{The performance comparison of MoDiffE and baseline models in three scenarios on three datasets.}
\label{tab:main_result}
\footnotesize
\begin{tabular}{ccccccc}
\hline \hline
\multicolumn{7}{c}{\textbf{Youshu}}                                                                                                                                                   \\ \hline 
\multicolumn{1}{c|}{\multirow{2}{*}{Model}} & \multicolumn{3}{c|}{Recall@20}                                  & \multicolumn{3}{c}{NDCG@20}                \\ \cline{2-7} 
\multicolumn{1}{c|}{}                                & Cold-Start   & All-Bundle    & \multicolumn{1}{c|}{Warm-Start}   & Cold-Start   & All-Bundle    & Warm-Start   \\ \hline
\multicolumn{1}{c|}{DropoutNet~\cite{item_cold_start_DropoutNet_2017}}             & 0.0022          & 0.0148          & \multicolumn{1}{c|}{0.0336}          & 0.0007          & 0.0055          & 0.0153          \\
\multicolumn{1}{c|}{CB2CF~\cite{item_cold_start_CB2CF_2019}}                  & 0.0012          & 0.0028          & \multicolumn{1}{c|}{0.0258}          & 0.0007          & 0.0021          & 0.0208          \\
\multicolumn{1}{c|}{Heater~\cite{item_cold_start_Heater_2020a}}                 & 0.0016          & 0.0541          & \multicolumn{1}{c|}{0.1753}          & 0.0007          & 0.0286          & 0.0826          \\
\multicolumn{1}{c|}{GAR-CF~\cite{item_cold_start_GAR_2022}}                 & 0.0015          & 0.0529          & \multicolumn{1}{c|}{0.1697}          & 0.0011          & 0.0317          & 0.0726          \\
\multicolumn{1}{c|}{GAR-GNN~\cite{item_cold_start_GAR_2022}}                & 0.0013          & 0.0358          & \multicolumn{1}{c|}{0.0835}          & 0.0006          & 0.0178          & 0.0569          \\
\multicolumn{1}{c|}{CVAR~\cite{item_cold_start_CVAR_2022}}                   & 0.0008          & 0.0829          & \multicolumn{1}{c|}{0.1958}          & 0.0002          & 0.0533          & 0.1112          \\
\multicolumn{1}{c|}{CLCRec~\cite{item_cold_start_CLCRec_2021}}                 & 0.0137          & 0.0367          & \multicolumn{1}{c|}{0.0626}          & 0.0087          & 0.0194          & 0.0317          \\
\multicolumn{1}{c|}{CCFCRec~\cite{item_cold_start_CCFCRec_2023}}                & 0.0044          & 0.0702          & \multicolumn{1}{c|}{0.1554}          & 0.0022          & 0.0425          & 0.0798          \\
\multicolumn{1}{c|}{CoHeat~\cite{bundle_cold_start_CoHeat_2024}}                 & \underline{0.0183}    & \underline{0.1247}    & \multicolumn{1}{c|}{\underline{0.2804}}    & \underline{0.0105}    & \underline{0.0833}    & \underline{0.1646}    \\ \hline
\multicolumn{1}{c|}{\textbf{MoDiffE}}                & \textbf{0.1210} & \textbf{0.1838} & \multicolumn{1}{c|}{\textbf{0.2846}} & \textbf{0.0890} & \textbf{0.1090} & \textbf{0.1697} \\ \hline \hline
\multicolumn{7}{c}{\textbf{NetEase}}                                                                                                                                                                          \\ \hline
\multicolumn{1}{c|}{\multirow{2}{*}{Model}} & \multicolumn{3}{c|}{Recall@20}                                              & \multicolumn{3}{c}{NDCG@20}                            \\ \cline{2-7} 
\multicolumn{1}{c|}{}                                & Cold-Start & All-Bundle & \multicolumn{1}{c|}{Warm-Start} & Cold-Start & All-Bundle & Warm-Start \\ \hline
\multicolumn{1}{c|}{DropoutNet~\cite{item_cold_start_DropoutNet_2017}}             & 0.0028              & 0.0046              & \multicolumn{1}{c|}{0.0154}              & 0.0015              & 0.0024              & 0.0078              \\
\multicolumn{1}{c|}{CB2CF~\cite{item_cold_start_CB2CF_2019}}                  & 0.0016              & 0.0027              & \multicolumn{1}{c|}{0.0049}              & 0.0006              & 0.0014              & 0.0027              \\
\multicolumn{1}{c|}{Heater~\cite{item_cold_start_Heater_2020a}}                 & 0.0021              & 0.0102              & \multicolumn{1}{c|}{0.0125}              & 0.0010              & 0.0054              & 0.0064              \\
\multicolumn{1}{c|}{GAR-CF~\cite{item_cold_start_GAR_2022}}                 & 0.0010              & 0.0014              & \multicolumn{1}{c|}{0.0063}              & 0.0005              & 0.0008              & 0.0035              \\
\multicolumn{1}{c|}{GAR-GNN~\cite{item_cold_start_GAR_2022}}                & 0.0009              & 0.0027              & \multicolumn{1}{c|}{0.0056}              & 0.0003              & 0.0012              & 0.0030              \\
\multicolumn{1}{c|}{CVAR~\cite{item_cold_start_CVAR_2022}}                   & 0.0002              & 0.0156              & \multicolumn{1}{c|}{0.0308}              & 0.0001              & 0.0084              & 0.0154              \\
\multicolumn{1}{c|}{CLCRec~\cite{item_cold_start_CLCRec_2021}}                 & 0.0136              & 0.0259              & \multicolumn{1}{c|}{0.0407}              & 0.0075              & 0.0138              & 0.0215              \\
\multicolumn{1}{c|}{CCFCRec~\cite{item_cold_start_CCFCRec_2023}}                & 0.0007              & 0.0130              & \multicolumn{1}{c|}{0.0265}              & 0.0004              & 0.0068              & 0.0128              \\
\multicolumn{1}{c|}{CoHeat~\cite{bundle_cold_start_CoHeat_2024}}                 & \underline{0.0191}        & \underline{0.0453}        & \multicolumn{1}{c|}{\underline{0.0847}}        & \underline{0.0093}        & \underline{0.0264}        & \underline{0.0455}        \\ \hline
\multicolumn{1}{c|}{\textbf{MoDiffE}}                & \textbf{0.0235}     & \textbf{0.0503}     & \multicolumn{1}{c|}{\textbf{0.0884}}     & \textbf{0.0125}     & \textbf{0.0285}     & \textbf{0.0476}     \\ \hline \hline
\multicolumn{7}{c}{\textbf{iFashion}}                                                                                                                                                                          \\ \hline
\multicolumn{1}{c|}{\multirow{2}{*}{Model}} & \multicolumn{3}{c|}{Recall@20}                                              & \multicolumn{3}{c}{NDCG@20}                            \\ \cline{2-7} 
\multicolumn{1}{c|}{}                                & Cold-Start & All-Bundle & \multicolumn{1}{c|}{Warm-Start} & Cold-Start & All-Bundle & Warm-Start \\ \hline
\multicolumn{1}{c|}{DropoutNet~\cite{item_cold_start_DropoutNet_2017}}             & 0.0009              & 0.0039              & \multicolumn{1}{c|}{0.0060}              & 0.0008              & 0.0027              & 0.0045              \\
\multicolumn{1}{c|}{CB2CF~\cite{item_cold_start_CB2CF_2019}}                  & 0.0009              & 0.0066              & \multicolumn{1}{c|}{0.0057}              & 0.0006              & 0.0048              & 0.0043              \\
\multicolumn{1}{c|}{Heater~\cite{item_cold_start_Heater_2020a}}                 & 0.0015              & 0.0123              & \multicolumn{1}{c|}{0.0217}              & 0.0010              & 0.0083              & 0.0151              \\
\multicolumn{1}{c|}{GAR-CF~\cite{GAN}}                 & 0.0013              & 0.0090              & \multicolumn{1}{c|}{0.0203}              & 0.0013              & 0.0055              & 0.0143              \\
\multicolumn{1}{c|}{GAR-GNN~\cite{GAN}}                & 0.0065              & 0.0126              & \multicolumn{1}{c|}{0.0172}              & 0.0030              & 0.0087              & 0.0107              \\
\multicolumn{1}{c|}{CVAR~\cite{item_cold_start_CVAR_2022}}                   & 0.0007              & 0.0125              & \multicolumn{1}{c|}{0.0220}              & 0.0004              & 0.0084              & 0.0152              \\
\multicolumn{1}{c|}{CLCRec~\cite{item_cold_start_CLCRec_2021}}                 & 0.0053              & 0.0126              & \multicolumn{1}{c|}{0.0203}              & 0.0043              & 0.0085              & 0.0135              \\
\multicolumn{1}{c|}{CCFCRec~\cite{item_cold_start_CCFCRec_2023}}                & 0.0005              & 0.0252              & \multicolumn{1}{c|}{0.0439}              & 0.0003              & 0.0172              & 0.0304              \\
\multicolumn{1}{c|}{CoHEAT~\cite{bundle_cold_start_CoHeat_2024}}                 & \underline{0.0170}        & \underline{0.0658}        & \multicolumn{1}{c|}{\underline{0.1156}}        & \underline{0.0096}        & \underline{0.0504}        & \underline{0.0876}        \\ \hline
\multicolumn{1}{c|}{\textbf{MoDiffE}}                & \textbf{0.0294}     & \textbf{0.0698}     & \multicolumn{1}{c|}{\textbf{0.1182}}     & \textbf{0.0184}     & \textbf{0.0540}     & \textbf{0.0893}     \\ \hline \hline
\end{tabular}
\end{table}

The performance comparisons of MoDiffE and baseline models in cold-start, all-bundle, and warm-start scenarios on Youshu, NetEase, and iFashion are shown in Table~\ref{tab:main_result}. 
The larger Recall@20 and NDCG@20, the better the performance. 
Based on the Table~\ref{tab:main_result}, we have the following main points: 
\begin{itemize}
    \item \textbf{MoDiffE achieves best performance in three scenarios of three datasets.} MoDiffE surpasses all the item cold-start recommendation baselines and the SOTA bundle cold-start recommendation model CoHeat~\cite{bundle_cold_start_CoHeat_2024} in all experiment settings. Especially in the cold-start and all-bundle scenarios, which are the primary targets of the bundle cold-start recommendation task, MoDiifE achieves significant improvements. 
    Specifically, in the Youshu dataset, MoDiffE achieves an absolute improvement of 0.1014 and 0.0778 in terms of Recall@20 and NDCG@20 in the cold-start scenario  and a relative improvement of 47.43\% and 30.94\% in the all-bundle scenario. 
    Even in the extreme warm-start case, i.e., iFashion’s warm-start scenario, MoDiffE performs comparably to the best baseline. 
    \item \textbf{Compared mainly focus on the bundle-level cold-start problem, MoDiffE, which solves dual-level cold-start problems, achieves better performance.} 
    The bundle cold-start problem is caused by the cold-start problems from different views at different levels, and there are differences between these sub-problems. 
    The existing item-based cold-start baselines cannot solve all of them. They focus on addressing the problem of poor representation of bundles that lack interaction while overlooking the fact that cold items exist in bundles, resulting in suboptimal recommendation performance. 
    MoDiffE recognizes the dual-level multi-view complexity of the bundle cold-start problem and uses diffusion models to solve all cold-start problems of each view at dual levels simultaneously, achieving the best results. 
    \item \textbf{Compared with CoHeat that emphasizes item-level information, MoDiifE, which fully uses dual-level information, achieves better performance.} 
    The iFashion dataset has filtered out most of the cold items, so CoHeat is least affected by the item-level cold-start problem in this dataset. 
    However, its Recall@20 and NDCG@20 of 0.0170 and 0.0096 in the cold-start scenario are still lower than MoDiffE’s 0.0294 and 0.0184.
    This demonstrates that bundle-level signals are crucial for bundle recommendation. 
    MoDiffE uses a diffusion model to generate bundle-level diffusion representations to supplement bundle-level embedded representations that are poorly represented due to missing features, thereby more effectively alleviating the bundle cold-start problem.

\end{itemize}

\subsection{Ablation Study}

Next, we investigate the effects of several key designs in MoDiffE through ablation studies on three datasets. 
As shown in Table~\ref{tab:ablation_study}, starting from MoDiifE, we remove modules one by one from top to bottom. The model variants used in ablation studies are as follows:
\begin{itemize}
    \item \textbf{\textit{w/o aug}}: removing the cold-start gating augmentation method (Section~\ref{sec:Cold-Start_Gating_Augmentation}) which interpolation-based synthesizes pseudo cold bundles during training of gating networks. This variant trains gating networks with the original training set. 
    \textbf{\item \textit{w/o moe}}: removing the cold-aware hierarchical MoE architecture, which combines results of cold-start sub-problems and adaptively gates different bundles with diverse cold situations (Section~\ref{sec:combine}). This variant adds up outputs of prior-embedding models and diffusion models as the final representation. 
    \textbf{\item \textit{w/o diff}}: removing all diffusion models (Section~\ref{sec:conquer}) and keep only prior-embedding models. This variant is the backbone model. 
\end{itemize}

The result of the ablation study is shown in Table~\ref{tab:ablation_study}. Here are our observations:
\begin{itemize}
    \item \textbf{ The gating networks in MoDiffE dependency on the cold-start gating augmentation method achieves coverage of cold bundles. } After removing the cold-start gating augmentation method, \textit{w/o aug} cannot perform cold-start bundle recommendation in the cold-start scenario, and Recall@20 and NDCG@20 are both 0. Since the bundle-level cold bundle is not present in the training set, gating networks in MoE are unable to handle it accurately during inference. 
    The performance of \textit{w/o aug} is close to \textit{w/o diff}, which is the backbone model, which shows that the potential of diffusion models and the cold-aware hierarchical MoE has not been stimulated, and MoDiffE has degenerated into the original backbone model. 
    \item \textbf{Exploiting the capabilities of all experts requires the effective combination of the cold-aware hierarchical MoE.}
    Compared with MoDiffE, \textit{w/o moe} removes the cold-aware hierarchical MoE, and its performance on all scenarios of the three datasets decreases overall. The decrease in Recall@20 and NDCG@20 ranges from 34.71\% to 96.27\%. 
    Moreover, \textit{w/o moe} is even lower than the backbone model \textit{w/o diff}, which explains that without a suitable combination, diffusion experts will not bring improvement and may even affect the performance of embedded experts in warm bundle recommendation. For example, in the all-bundle scenario on the Youshu dataset, the Recall@20 of \textit{w/o moe} is 0.0633 compared to 0.0876 of \textit{w/o diff}. 
    \item \textbf{Diffusion models offer essential capabilities to handle bundle cold-start problems. } The metrics of \textit{w/o diff} in the cold-start scenario in all datasets is 0. This is because the backbone model based on prior-embedding models cannot handle the cold-start problem caused by the lack of specific feature input. In contrast, diffusion models overcome this shortcoming by generating representations that are not dependent on specific features. 
\end{itemize}

\begin{table}[t]
\caption{The ablation study of MoDiffE in cold-start and all-bundle scenarios which are primary targets of cold-start bundle recommendation task. Cold and All are the cold-start scenario and the all-bundle scenario. }
\label{tab:ablation_study}
\scriptsize
\begin{tabular}{c|cccc|cccc|cccc}
\hline
Dataset                & \multicolumn{4}{c|}{\textbf{Youshu}}                                                 & \multicolumn{4}{c|}{\textbf{NetEase}}                                                & \multicolumn{4}{c}{\textbf{iFashion}}                                               \\ \hline
\multirow{2}{*}{Model} & \multicolumn{2}{c|}{Recall@20}      & \multicolumn{2}{c|}{NDCG@20} & \multicolumn{2}{c|}{Recall@20}      & \multicolumn{2}{c|}{NDCG@20} & \multicolumn{2}{c|}{Recall@20}      & \multicolumn{2}{c}{NDCG@20} \\ \cline{2-13} 
                                & Cold & \multicolumn{1}{c|}{All} & Cold        & All        & Cold & \multicolumn{1}{c|}{All} & Cold        & All        & Cold & \multicolumn{1}{c|}{All} & Cold        & All       \\ \hline
\textbf{MoDiffE }                        & \textbf{0.1210}   & \multicolumn{1}{c|}{\textbf{0.1838}}   & \textbf{0.0890}          & \textbf{0.1090}          & \textbf{0.0235}   & \multicolumn{1}{c|}{\textbf{0.0503}}   & \textbf{0.0125}          & \textbf{0.0285}          & \textbf{0.0294}   & \multicolumn{1}{c|}{\textbf{0.0698}}   & \textbf{0.0184}          & \textbf{0.0540}         \\ \hline
\textit{w/o aug}              & 0          & \multicolumn{1}{c|}{0.1294}   & 0                 & 0.0878          & 0          & \multicolumn{1}{c|}{0.0470}    & 0                 & 0.0273         & 0          & \multicolumn{1}{c|}{0.0609}   & 0                 & 0.0471         \\
\textit{w/o moe}           & 0.0100   & \multicolumn{1}{c|}{0.1001}   & 0.0045          & 0.0633          & 0.0012   & \multicolumn{1}{c|}{0.0328}   & 0.0004          & 0.0177          & 0.0015   & \multicolumn{1}{c|}{0.0415}   & 0.0008          & 0.0314         \\
\textit{w/o diff}            & 0          & \multicolumn{1}{c|}{0.1290}      & 0                 & 0.0876          & 0          & \multicolumn{1}{c|}{0.0461}   & 0                 & 0.0266          & 0          & \multicolumn{1}{c|}{0.0658}   & 0                 & 0.0518  \\ \hline
\end{tabular}
\end{table}

\subsection{Hyperparameter Experiments}
We report the effects of different hyperparameters of several modules to facilitate future applications of MoDiffE. Our experiments are conducted on Youshu and NetEase datasets because they conduct less sparsity filtering. We focus on the cold-start scenario and all-bundle scenario, which are the primary targets of the bundle cold-start recommendation task.

\subsubsection{Hyperparameters of Diffusion Model}

\begin{figure}[h]
\includegraphics[width=\textwidth]{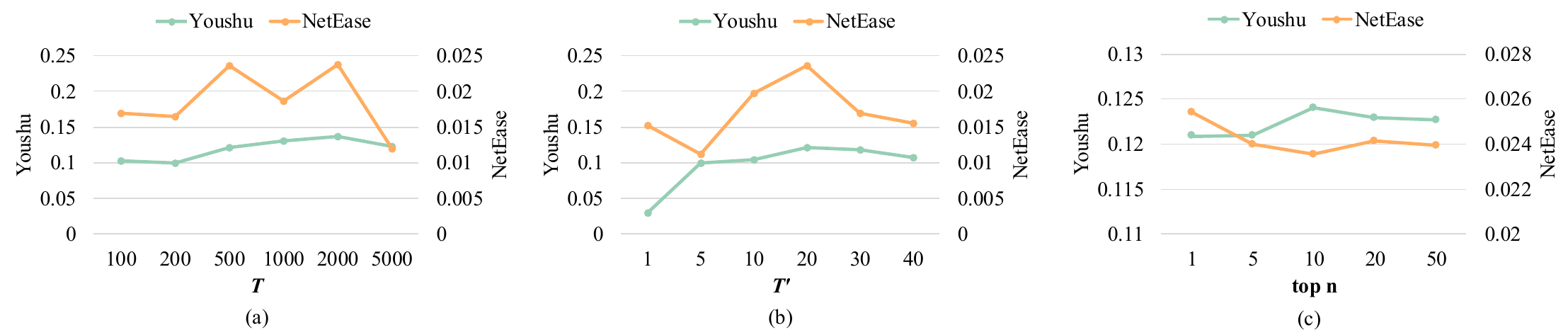}
\caption{Impact of diffusion model hyperparameters on Recall@20 in cold-start scenario on Youshu and NetEase: (a) the number of training steps $T$; (b) the number of inference steps $T'$; (c) top n of similarity-based sampling. }
\label{fig:p_diff}
\end{figure}

The number of training steps $T$ and the number of inference steps $T'$ are two important hyperparameters of diffusion models. 
Their experimental results are shown in Figure~\ref{fig:p_diff} (a) and (b). They follow similar trends. Too many or too few steps is not good for the performance of the model. 
In particular, too many of inference steps will also lead to slower inference speed. 

In Section~\ref{inference}, the similarity-based sampling method is used to improve the stability and performance of the representation generation. 
The impact of selecting different numbers of bundles is shown in Figure~\ref{fig:p_diff} (c).
A larger value of $n$ leads to a smoother initial representation for inference denoising. Youshu performs best when $n=10$, while NetEase achieves optimal results by using the representation of the most similar bundle as the initial input. Data analysis reveals that bundle similarity in NetEase is higher than in Youshu, which facilitates more accurate initial representations and benefits the denoising process during inference.

\subsubsection{Hyperparameters of Cold-Aware Hierarchical MoE}
MoDiffE adopts a cold-aware hierarchical MoE to combine the results of all sub-problems into the final recommendation, involving multiple gating networks, first performing view-level gating to fuse heterogeneous expert representations, and then output-level gating to fuse predictions from different views. 
The activation function of the gating network is a critical hyperparameter. We evaluate different activation combinations in two scenarios on two datasets, with results summarized in Table~\ref{tab:p_func}. 
Using ReLU or Tanh in both layers yields suboptimal performance. 
In contrast to the vanilla MoE, which typically employs Softmax as the activation function, MoDiffE adopts a combination of Softmax and Tanh to achieve superior results. 
This is because different views are not competitive but relatively cooperative, and Tanh better captures their collaborative relationship, leading to improved results.

\begin{table}[t]
\caption{Performance comparison w.r.t. selection of activation functions for different layers in cold-aware hierarchical MoE in terms of Recall@20.}
\label{tab:p_func}
\footnotesize
\begin{tabular}{cc|cc|cc}
\hline \hline
\multicolumn{2}{c|}{Activation Function} & \multicolumn{2}{c|}{Cold-Start Scenario} & \multicolumn{2}{c}{All-Bundle Scenario} \\ \hline
View Layer            & Output Layer       & Youshu              & NetEase            & Youshu             & NetEase            \\ \hline
Relu                  & Relu               & 0.0104             & 0.0015            & 0.1261            & 0.0435            \\
Tanh                  & Tanh               & 0.0143             & 0.0015            & 0.1056            & 0.0359            \\
Softmax               & Softmax            & 0.1171             & 0.0205            & 0.1784            & 0.0479            \\ \hline
\textbf{Softmax}      & \textbf{Tanh}      & \textbf{0.1210}    & \textbf{0.0235}   & \textbf{0.1838}   & \textbf{0.0503}   \\ \hline \hline
\end{tabular}
\end{table}

\begin{figure}[b]
\includegraphics[width=0.7\textwidth]{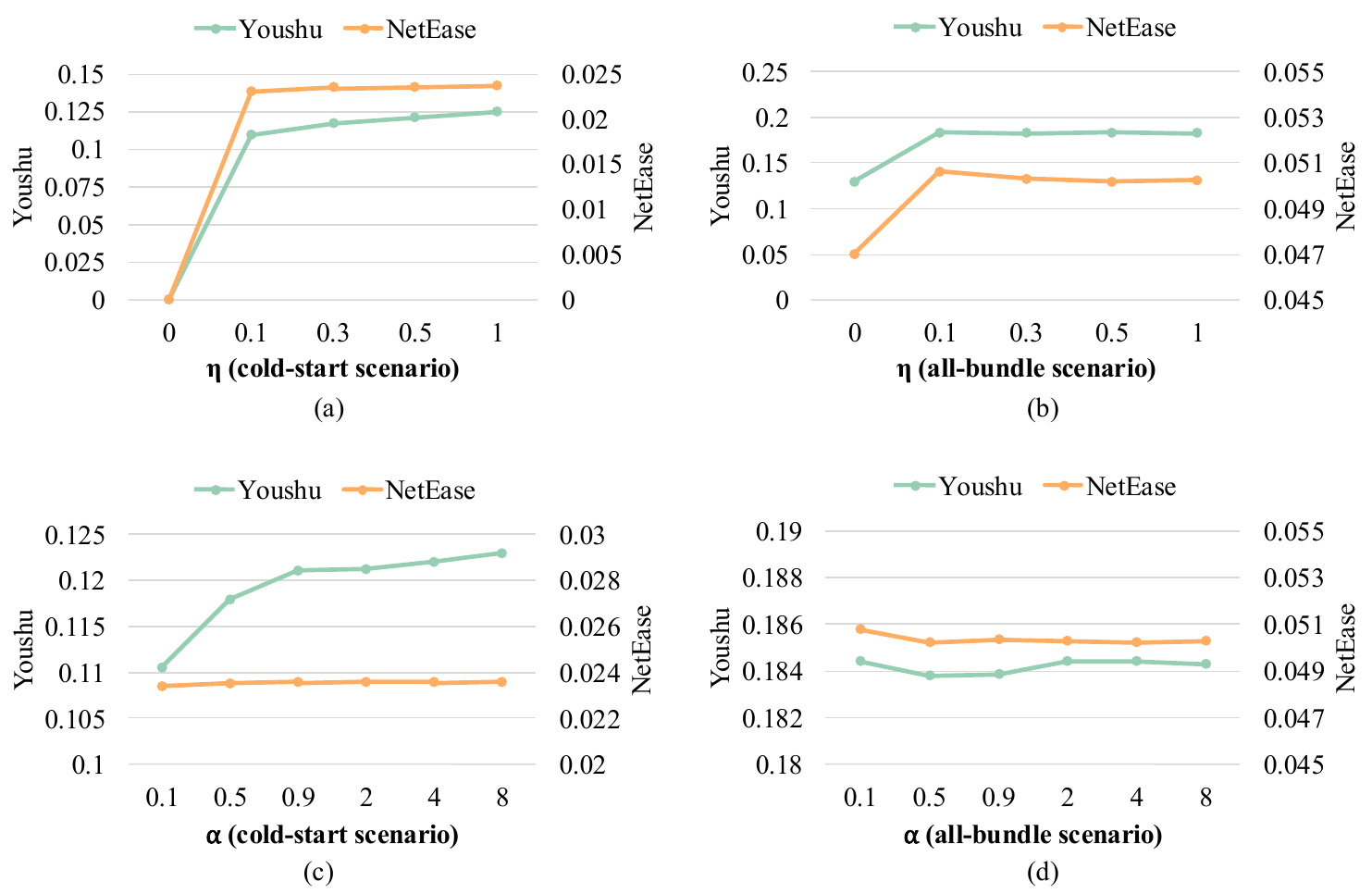}
\caption{Impact of the cold-start gating augmentation method's hyperparameters on Recall@20 in cold-start and all-bundle scenarios on Youshu and NetEase. (a) and (b): the interpolation sampling ratio $\eta$; (c) and (d): Beta$(\alpha,\alpha)$ distribution for sample interpolation. }
\label{fig:p_aug}
\end{figure}

\subsubsection{Hyperparameters of Cold-Start Gating Augmentation}
The interpolation sampling ratio $\eta$ and Beta$(\alpha,\alpha)$ distribution for sample interpolation are two hyperparameters of cold-start gating augmentation. 
The impact of different values on MoDiffE in two key scenarios of two datasets is shown in Figure~\ref{fig:p_aug}. 
The huge improvement of $\eta$ from 0 to 0.1 in (a) and (b) proves that the cold-start gating augmentation is crucial to handling cold entities. 
Further increasing $\eta$ can slightly improve the performance of cold-start scenarios at two datasets, but it will greatly extend the training time. Increasing $\eta$ in the all-bundle scenario does not bring more improvement. 
From Figure~\ref{fig:p_aug} (c) and (d), we can see that MoDiffE is insensitive to the $\alpha$ parameter in most cases and is only affected in the cold-start scenario of Youshu.

\subsection{Visualization Analysis of Representations}

\begin{figure}[h]
\includegraphics[width=0.9\textwidth]{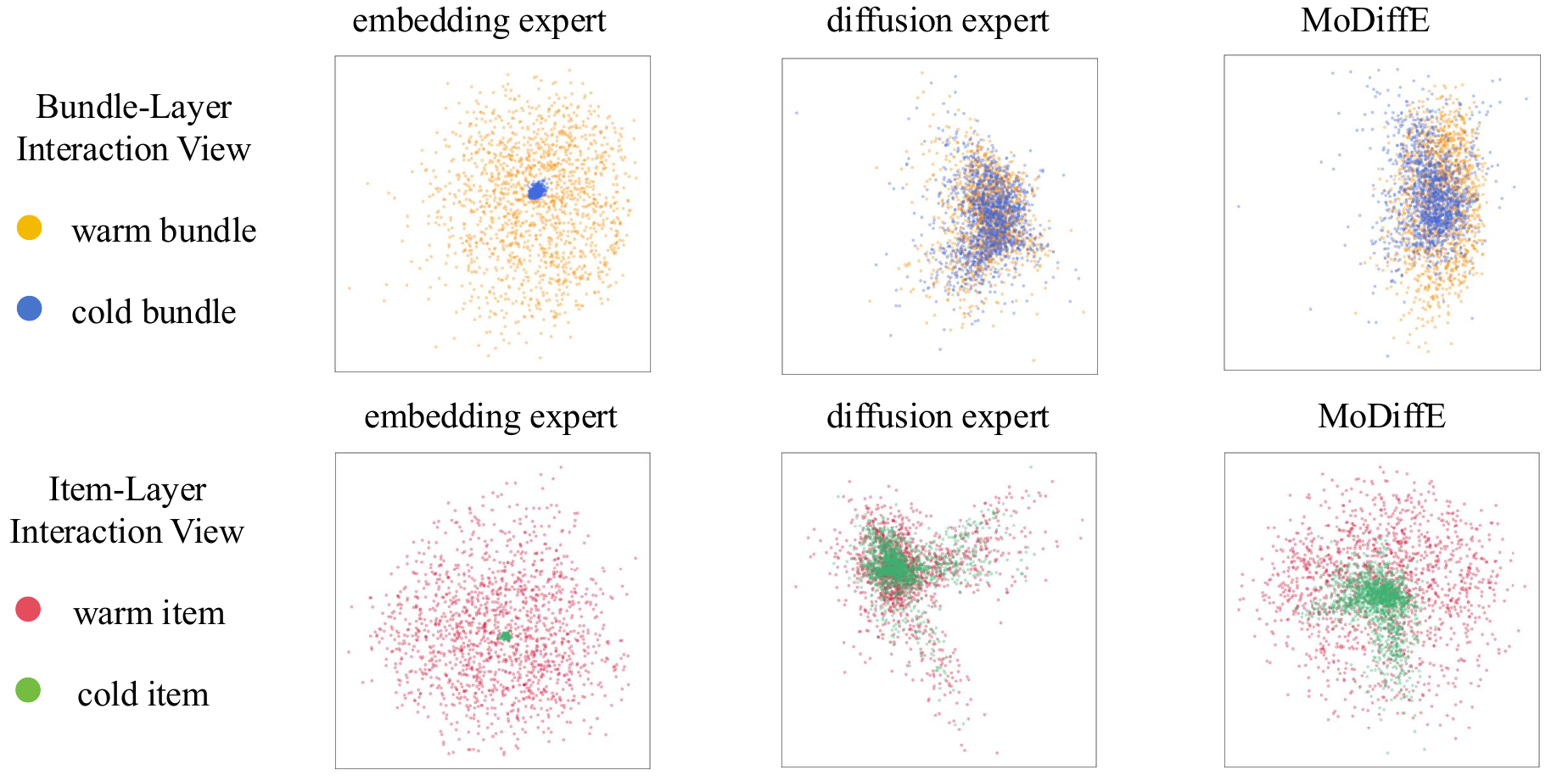}
\caption{Distributions of bundles/items in interaction views under different experts in the cold-start scenario of the NetEase dataset.}
\label{fig:distribuition}
\end{figure}

We visualize the learned bundle and item representations using the PCA algorithm to demonstrate that our MoDiffE effectively addresses the poor representation problem for feature-missing bundles and items in prior-embedding models.
Figure~\ref{fig:distribuition} illustrates the distribution of bundles and items in the interaction view under different experts in the cold-start scenario of the NetEase dataset.
Specifically, in embedding experts, cold bundles and items suffer from a lack of feature inputs, causing their representations to remain untrained and collapse around the initialization point, making them indistinguishable.
In contrast, diffusion experts do not rely on specific feature inputs; instead, they learn to generate representations by conducting diffusion-denoising training based on warm bundle/item representations.
As a result, both cold and warm bundles/items are more evenly distributed across the representation space.
Finally, the view-layer gating networks in MoDiffE adaptively fuse the outputs of the embedding experts and diffusion experts to form the final bundle and item representations. 
In contrast to the prior-embedding model, MoDiffE produces representations where cold bundles and items are no longer clustered together but are instead more evenly and coherently distributed with warm bundles and items in the representation space.

\subsection{Gating Case Study}
\begin{figure}[h]
\includegraphics[width=\textwidth]{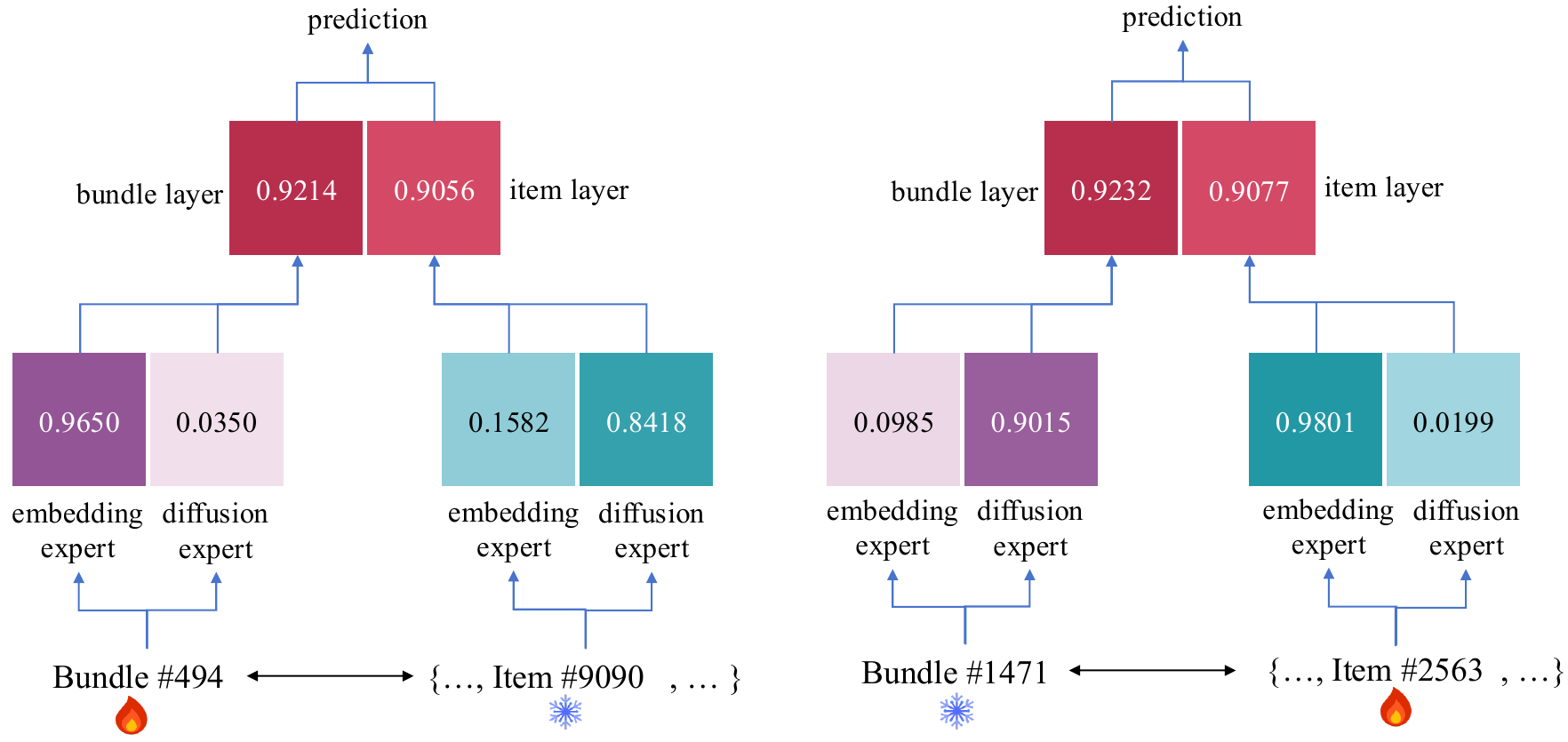}
\caption{View-layer gating and output-layer gating of warm bundle \#494, cold bundle \#1471, cold item \#9090, and warm item \#2563 in the all-bundle scenario of Youshu.}
\label{fig:gating}
\end{figure}

After visualizing representation fusion at the distribution level, we conduct a gating case study to specifically demonstrate how our cold-aware hierarchical MoE gating bundles/items in different situations. 
Figure~\ref{fig:gating} shows view-layer gating and output-layer gating of two bundles and two items in the all-bundle scenario of Youshu. 
For warm bundle \#494, its view-layer gating mainly leans on its own well-trained representation in the embedding expert, whereas for cold bundle \#1471 it relies mainly on the diffusion expert. 
The same gating pattern appears at the item layer: warm items lean to the embedding expert, and cold items to the diffusion expert. This avoids poor representations of embedding experts when features are missing. 
Figure~\ref{fig:gating} confirms that our MoDiffE can adaptively handle diverse cold-start situations of bundles. 

Existing bundle recommendation models~\cite{BGCN,CrossCBR,MIDGN,MultiCBR,HyperMBR,AMCBR} implicitly learn view cooperation and ultimately reflect in the representation
scales.
MoDiffE uses the diffusion model to improve the representation capabilities of different views but also breaks the original cooperative relationship between views. 
Our MoDiffE uses the output-layer gating network with Tanh activation function to relearn the new weight relationship instead of using equal weights. 
The output-layer gating in Figure~\ref{fig:gating}  shows that the current MoDiffE framework has become slightly biased towards the bundle-layer view. 
This is because the performance of the bundle recommendation model mainly depends on the bundle layer~\cite{BGCN,CrossCBR,MultiCBR}. After addressing the cold-start problem of the bundle layer, the model improves the utilization of bundle-layer views.

\subsection{Bundle Hit Analysis}
\begin{figure}[h]
\includegraphics[width=\textwidth]{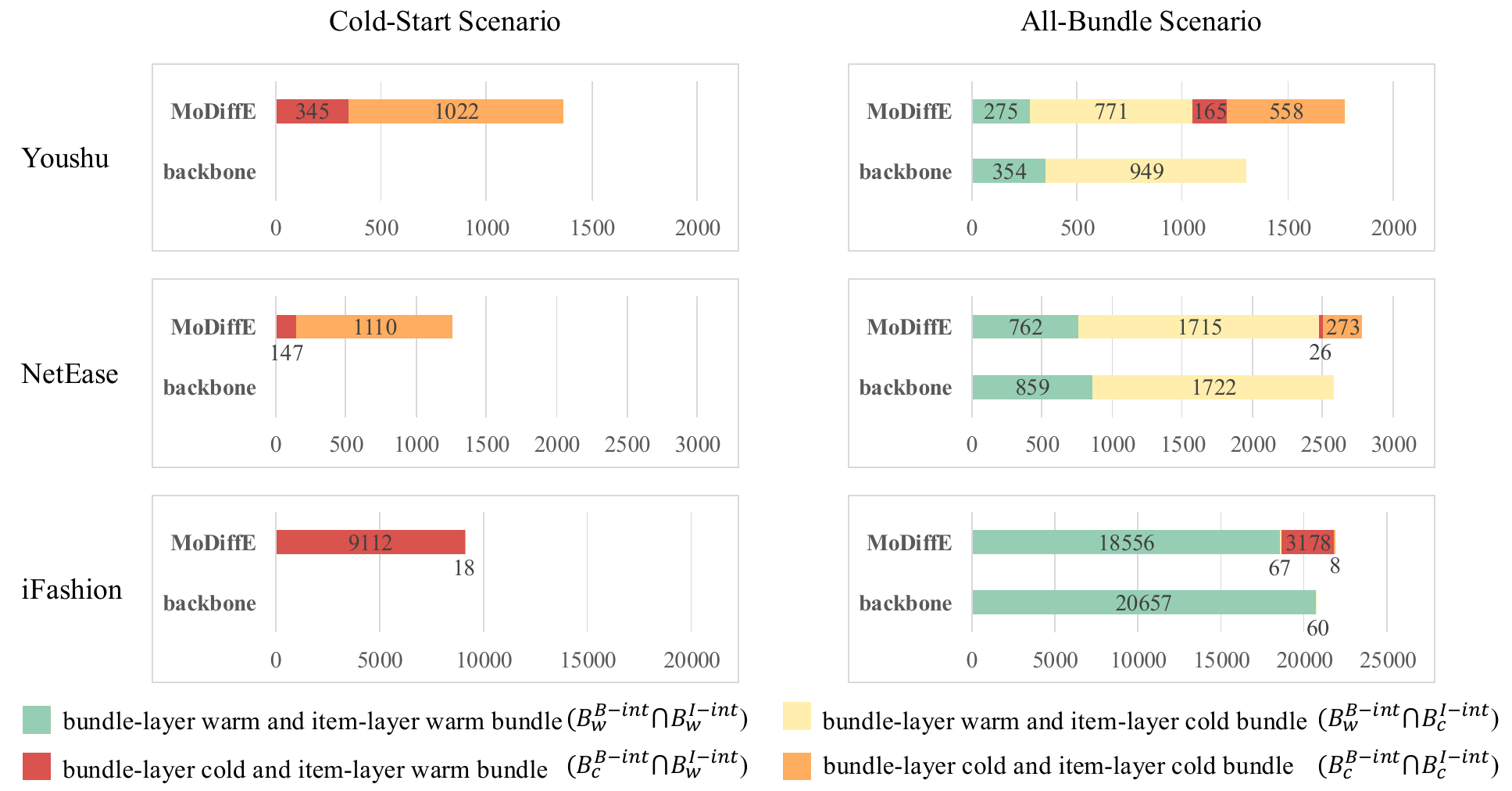}
\caption{Hit counts of bundles in different situations in the recommended top 20 lists of MoDiffE and backbone model in the cold-start and all-bundle scenarios of three datasets.}
\label{fig:hit}
\end{figure}

To evaluate MoDiffE’s recommendation of bundles in the different cold-start situations, we compare the top-20 hit count of MoDiffE versus the backbone model (CrossCBR~\cite{CrossCBR}) in both cold-start and all-bundle scenarios. 
Figure \ref{fig:hit} presents these results, note that the proportion of hitted bundles in different situations is related to the proportion in the original dataset (see Table~\ref{tab:data_analysis} for details). 
In the cold-start setting, the backbone model fails to recommend bundles lacking features, whereas our MoDiffE employs a divide-and-conquer strategy and uses diffusion models and the cold-aware hierarchical MoE to achieve recommendations for cold bundles. 
In the all-bundle scenario, the MoDiffE model also greatly improves the recommendation performance of cold bundles. However, due to the limited resources of the recommendation list, there is a seesaw phenomenon between bundles in different situations. The performance of the warm-start bundle is reduced, but the overall performance is improved, and the fairness of the cold-start bundle is also improved.

\section{Conclusion and Future Work}
The bundle cold-start problem is critical for bundle recommendation because it can improve user experience, increase merchant revenue, and improve bundle fairness. 
In this work, we are the first to emphasize the dual-level multi-view complexity of the bundle cold-start problem, which poses challenges in simultaneously addressing cold-start problems from multiple views at dual levels and adaptively handling diverse cold-start situations of bundles. 
To address the two challenges, we propose the MoDiffE framework, which employs a divide-and-conquer strategy. This framework divide the overall problem into similar sub-problems, leverages diffusion models to uniformly conquer different cold-start sub-problems, and uses a cold-aware hierarchical MoE to combine sub-problems and adaptively handle different bundles. MoDiffE adopts a multi-stage decoupled pipeline for training, which enhances the framework's flexibility and adaptability. Moreover, MoDiffE uses a cold-start gating augmentation method to achieve gating on cold bundles. 
Extensive experiments on three real-world datasets demonstrate that MoDiffE significantly outperforms state-of-the-art baselines in all cold-start, all-bundle, and warm-start scenarios. 
The ablation experiment proves the necessary contribution of different components to the framework. The visual analysis shows the representation learning of different experts and MoDiffE's fusion effects from a distribution perspective. 
Gating case study and bundle hit analysis specifically show the modeling and recommendation performance of the model for different situations of cold start bundles. 

Moving forward, there are two directions worth exploring. 
First, our framework currently uses the uni-modal to uni-modal diffusion model to generate feature representations, which cannot fully utilize the multi-modal information of the bundle. We consider introducing the multi-modal diffusion model~\cite{MM_diffusion_2023, MM_diffusion_2024} to achieve more efficient and robust generation. 
Second, we will attempt to integrate the user cold-start problem into the same divide-and-conquer paradigm will provide a comprehensive solution to the cold-start problem in the field of bundle recommendations. 
We expect that MoDiffE provides useful insights for future research on multi-view cold-start recommendations and encourages further exploration of inclusive and robust recommender systems.

\begin{acks}
This research is partially supported by NSFC, China (No.62276196). This research is also supported by the Natural Sciences and Engineering Research Council (NSERC) of Canada and the York Research Chairs (YRC) program.
\end{acks}

\bibliographystyle{ACM-Reference-Format}
\bibliography{references}

\end{document}